\documentclass[authoryear,12pt,3p]{elsarticle}

\usepackage[utf8]{inputenc}
\usepackage[T1]{fontenc}
\usepackage[UKenglish]{babel}

\usepackage{tabularx}
\usepackage{xcolor}
\usepackage{graphicx}
\graphicspath{{images/}}

\usepackage{multirow}

\usepackage{numprint}
\usepackage[squaren,Gray]{SIunits}

\usepackage{hyperref}

\usepackage{wasysym}

\usepackage{textcomp} % For \textdegree

\begin{document}

\title{SPH calculations of Mars-scale collisions: the role of the Equation of State, material rheologies, and numerical effects}
\author[unibe]{Alexandre~Emsenhuber\corref{cor}}
\ead{alexandre.emsenhuber@space.unibe.ch}
\author[unibe]{Martin~Jutzi}
\ead{martin.jutzi@space.unibe.ch}
\author[unibe]{Willy~Benz}
\ead{willy.benz@space.unibe.ch}
\cortext[cor]{Corresponding author}
\address[unibe]{Physikalisches Institut \& Center for Space and Habitability, Universität Bern, Gesellschaftsstrasse 6, CH-3012 Bern, Switzerland}

\begin{abstract}
We model large-scale ($\approx\unit{2000}{\kilo\meter}$) impacts on a Mars-like planet using a Smoothed Particle Hydrodynamics code. The effects of material strength and of using different Equations of State on the post-impact material and temperature distributions are investigated. The properties of the ejected material in terms of escaping and disc mass are analysed as well. We also study potential numerical effects in the context of density discontinuities and rigid body rotation.
We find that in the large-scale collision regime considered here (with impact velocities of $\unit{4}{\kilo\meter\per\second}$), the effect of material strength is substantial for the post-impact distribution of the temperature and the impactor material, while the influence of the Equation of State is more subtle and present only at very high temperatures.  
\end{abstract}


\maketitle
\thispagestyle{plain}

% --------------------
\section{Introduction}
% --------------------

Giant impacts occurring at the end stages of planet formation define the properties of the final planets and moons. Examples include Mercury’s anomalously thin silicate mantle \citep{1988IcarusBenzSlatteryCameron,2007SSRvBenzAnicHornerWhitby,2014AGUFMAsphaugReufer}, the origin of the Earth’s Moon \citep{1975IcarusHartmannDavis,1976LPICameronWard,2001NatureCanupAsphaug,2004IcarusCanup,2012ScienceCanup,2012ScienceCukStewart,2012IcarusReufer} or the formation of the Pluto-Charon system \citep{2005ScienceCanup,2011AJCanup}. Smaller, but still planet-scale collisions were proposed for the formation of the martian dichotomy \citep{1984NatureWilhelmsSquyres,1988GeoRLFreySchultz,2008NatureAndrewsHanna,2008NatureNimmo,2008NatureMarinova,2011IcarusMarinova}.
Numerical simulations of such giant impact scenarios are performed using shock physics codes, which are often based on the smoothed particle hydrodynamics (SPH) method. In such simulations of planet-scale collisions it is typically assumed that material strength is negligible due to large overburden pressures caused by self-gravitation. However, it was found in previous studies \citep[e.g.][]{2015PSSJutzi} that for collisions at $\unit{100}{\kilo\meter}$ to $\unit{1000}{\kilo\meter}$ scale, the effect of material strength can still be very important.
The impactor size range for which the assumption of purely hydrodynamic behaviour is justified has not yet been studied systematically.
Other uncertainties in giant impact simulations are related to the Equation of State (EoS) models. Often used EoS are the simple Tillotson \citep{Tillotson} EoS \citep[e.g.][]{2008NatureMarinova,2011IcarusMarinova} or the more sophisticated (M-)ANEOS \citep{ANEOS,2007M&PSMelosh}. Last but not least, there are potential numerical issues that have to be added to the uncertainties. Concerning the SPH method, known numerical issues are related to surface and contact discontinuities \citep[e.g.][]{2013PASJHosonoSaitohMakino,2017MNRASReinhardt} or rotational instabilities in rigid body rotations \citep[e.g.][]{Habil2006Speith}.

In this study, we focus on large-scale ($\approx\unit{2000}{\kilo\meter}$) impacts on a Mars-like planet. The conditions studied here are chosen to be quite general, but at the same time they also cover a range of possible dichotomy forming events \citep{2008NatureMarinova}. We want to assess the relative importance of the effects mentioned above (i.e., material strength; choice of EoS; numerical issues) in collision simulations of such scale. This knowledge is important in order be able to make an educated choice of the material models and numerical methods, and to know their limitations.

Collisions are modelled using an updated SPH code \citep{2012IcarusReufer}, which includes self-gravity, a newly implemented strength model \citep[following][]{2015PSSJutzi} and various EoS (Tilloston and M-ANEOS). We investigate the effects of the above mentioned material properties (namely material strength and the EoS models) on the outcome of the collisions for different impact geometries. We focus on the material and temperature distributions in the final planet but also analyse the properties of the ejected material (orbiting and unbound). Finally, potential numerical effects are studied as well. We investigate the known SPH issue in the case of rigid body rotation. We also consider different schemes to compute the density, in order to investigate potential numerical issues at discontinuities (such as the free surface and the core-mantle boundary).

A subset of the SPH calculations presented here is coupled with thermochemical simulations of planetary interiors in a companion paper \citep{Companion}. The effects of the different models on the long-term evolution (over a time period of about $\unit{0.5}{\mega yr}$) and the crust formation will be discussed there.

In section 2 we describe our modeling approach and discuss in detail the implemented strength model and its coupling with the EoS models. The initial conditions (target and projectile properties and impact geometry) are detailed in section 3. In section 4, we present the results of our simulations using different material models and numerical schemes, followed by a discussion and conclusions in section 5.

% --------------------------
\section{Modelling approach}\label{sec:modeling}
% --------------------------

We use a smoothed particle hydrodynamics (SPH) code \citep[e.g.][]{1994IcarusBenzAsphaug,2012IcarusReufer,2015PSSJutzi} to model the impact event.  The code used here is based on the SPHLATCH code developed by \citet{2012IcarusReufer}. It includes a newly implemented pressure and temperature dependent shear strength model, as described below, which is appropriate for the large scale collisions considered here.
To study smaller scale collisions in geological materials, we use a different code, which also includes a tensile fracture and a porosity model \citep[see][for details]{2015PSSJutzi}. 

SPH uses a Lagrangian representation where material is divided into particles. Quantities are interpolated (‘smoothed’) by summing over surrounding particles (called neighbours) according to
\begin{equation}
B(\vec{x})=\sum_iB_iW(\vec{x}-\vec{x_i},h_i)V_i
\label{eq:sphsum}
\end{equation}
where $B_i$ represents the quantity (field variable) to be interpolated, and $\vec{x}_i$, $h_i$ and $V_i$ are the position, smoothing length, and volume of particle i, respectively. $W(\vec{x},h)$ is a smoothing kernel, which has the propriety to vanish if $\|\vec{x}\| > 2h$, so that the hydrodynamic variables (pressure, density, stress tensor) are integrated over a local group of neighbouring particles. In our simulations, we use a 3D cubic spline kernel \citep{1985A&AMonaghanLattanzio}. 

This interpolation scheme is used in SPH to solve the relevant differential equations. In standard SPH, the density is computed by direct summation (equation \ref{eq:sphsum}). We refer to this method as \textit{density summation} through this article. Alternatively, it can be computed by using the changing rate of density, which is given by the continuity equation:
\begin{equation}
\frac{D\rho}{Dt}+\rho\frac{\partial v_x}{\partial x}+\rho\frac{\partial v_y}{\partial y}+\rho\frac{\partial v_z}{\partial z}=0
\label{eq:model:densint}
\end{equation}
where $D/Dt$ stands for the substantive time derivative. We call it \textit{density integration}. A similar equation gives the time derivative of the smoothing length:
\begin{equation}
\frac{Dh}{Dt}+h\frac{\partial v_x}{\partial x}+h\frac{\partial v_y}{\partial y}+h\frac{\partial v_z}{\partial z}=0
\end{equation}
Thus the smoothing length grows in sparse regions and shrinks in dense regions, so that the number of neighbours within $2h$ is $\approx 50$. 

\begin{table}
	\begin{center}
		\begin{tabular}{ccccc}
			\hline
			& \multicolumn{2}{c}{ANEOS}  & \multicolumn{2}{c}{Tillotson} \\
			& iron & silicate & iron & olivine \\
			\hline
			G [GPa] & 76 & 44.4 & 76 & 72 \\
			$\sigma_M$ [GPa] & 0.68 & 3.5 & 0.68 & 3.5 \\
			$u_m$ [J/kg] & - & - & $9.2\cdot10^5$ & $3.4\cdot10^6$ \\
			$T_m$ [K] & - & - & 920 & 3400 \\
			\hline
		\end{tabular}
		
		\caption{Material parameters for the solid rheology model. G is the shear modulus. $\sigma_M$ is the von Mises plastic limit. Melting temperature $T_m$ is computed from $u_m$ using equation (\ref{eq:model:tillotsontemp}).}
		\label{tab:solid}
	\end{center}
\end{table}

Body accelerations are the result of the pressure gradient, and for this we use the pressure as computed from an equation of state (EoS) (see below), which is a function $P(\rho,u)$ of density $\rho$ (see equation \ref{eq:model:densint}) and internal energy $u$. For solid materials the pressure gradient is generalised into a stress tensor \citep{1994IcarusBenzAsphaug} with the resulting equation of motion given by: 
\begin{equation}
\frac{Dv_a}{Dt}=\frac{1}{\rho}\frac{\partial \tau_{ab}}{\partial x_b}
\label{eq:model:velmat}
\end{equation}
The stress tensor is defined as
\begin{equation}
\tau_{ab}=-P\delta_{ab}+\sigma_{ab}
\label{eq:model:tau}
\end{equation}
where $P$ is the pressure, $\delta_{ab}$ is the Kronecker symbol and $\sigma_{ab}$ is the traceless deviatoric stress tensor. The time evolution of $\sigma_{ab}$ is computed adopting Hooke’s law as in \citet{1994IcarusBenzAsphaug,1995IcarusBenzAsphaug}. In the limit of zero deviatoric stress this reduces to the familiar pressure-gradient acceleration. Energy conservation now reads
\begin{equation}
\frac{Du}{Dt}=\frac{1}{\rho}\tau_{\alpha\beta}\dot{\varepsilon}_{\alpha\beta}
\label{eq:model:specenergy}
\end{equation}
with $\dot{\varepsilon}_{\alpha\beta}$ being the strain rate tensor.

In standard SPH, the computation of the strain and rotation rate tensors fails to conserve angular momentum in rigid body rotations. Following the approach by \citet{Habil2006Speith}, we use a correction tensor in the computation of the velocity derivatives, which allows conservation of angular momentum at the cost of an additional computation step.

Equations (\ref{eq:model:velmat}) and (\ref{eq:model:tau}) describe an entirely elastic material. To model plastic behaviour we use a Drucker-Prager-like yield criterion \citep[e.g.][]{2004M&PSCollins,2015PSSJutzi}. The model has one yield strength $\sigma_i$, corresponding to intact material, and another $\sigma_d$ corresponding to completely fragmented material:
\begin{eqnarray}
\sigma_i&=&C+\frac{\mu_iP}{1+\mu_iP/(\sigma_M-C)} \\
\sigma_d&=&\mu_dP
\end{eqnarray}
where $C$ is the cohesion (yield strength as zero pressure), $\sigma_M$ is the von Mises plastic limit, and $\mu_i$ and $\mu_d$ are the coefficients of friction for intact and completely damaged material, respectively. We set them as $\mu_i=2$ and $\mu_d=0.8$ as in \citet{2004M&PSCollins}.  Note that $\sigma_d$ is limited to $\sigma_d<\sigma_i$ at high pressures. For the large-scale collisions considered here cohesion and tensile strength are negligible due to the large gravity-induced lithostatic stress (see e.g. Jutzi et al., 2015). We therefore set $C=0$. However, shear strength (limited by $\sigma_M$) is still important and cannot be neglected, as we shall see below.

The yield strength of intact material is a function of temperature. To take this into account, we adopt the same relation as in \citet{2004M&PSCollins}:
\begin{equation}
\sigma \to \sigma\tanh{\left(\xi\left(\frac{T_m}{T}-1\right)\right)}
\label{eq:model:tanh}
\end{equation}
where $\xi$ is a material constant which we set to $\xi=1.2$, $T$ is the temperature and $T_m$ is the corresponding liquidus temperature (see below). If $T>T_m$ the material is molten and we set $\sigma=0$. Finally, if the measure of the stress state, the second invariant of the deviatoric stress tensor
\begin{equation}
\sigma_{ll}=\sqrt{\frac{1}{2}\sigma_{ab}\sigma_{ab}}
\end{equation}
exceeds $\sigma$, the components of the deviatoric stress tensor are reduced by a factor $\sigma/\sigma_{ll}$. The parameters for the solid rheology model are provided in table~\ref{tab:solid}.

To complete the set of equations, an equation of state providing a relation between pressure, temperature and density is needed. We use either Tillotson \citep{Tillotson} or (M-)ANEOS \citep{ANEOS,2007M&PSMelosh}. The Tillotson equation of state provides both the pressure $P$ and the speed of sound as output, but it does not give the temperature directly, so as an approximation we use the internal energy as a proxy for temperature by dividing by the heat capacity
\begin{equation}
T=u/c_p
\label{eq:model:tillotsontemp}
\end{equation}
In this case equation (\ref{eq:model:tanh}) is modified to use $u$ and $u_m$ instead of $T$ and $T_m$ respectively, and $u_m$ is treated as a material constant independent of pressure. Iron parameters are from \citet{Tillotson} \citep[also in][]{1989BookMelosh} and the ones for olivine from \citet{2008NatureMarinova}.

\begin{table}
	\begin{center}
		\begin{tabular}{lll}
			\hline
			Name  & Description & Value \\
			\hline
			$\rho_0$ & Reference density & $\unit{7850}{\kilo\gram\per\meter\cubic{}}$ \\
			$T_0$ & Reference temperature & $0$ \\
			$p_0$ & Pressure at the reference point & $0$ \\
			$B_0$ & Bulk modulus at the reference point & $\unit{1.45\cdot10^{11}}{\pascal}$ \\
			$T_\mathrm{Debye}$ & Reference Debye temperature & $\unit{464}{\kelvin}$ \\
			$\gamma_\mathrm{max}$ & Limiting value of the Gruneisen coefficient & $2/3$ \\
			$E_\mathrm{sep}$ & Zero temperature separation energy & $\unit{2\cdot10^{8}}{\joule\per\kilo\gram}$ \\ 
			$T_\mathrm{melt}$ & Melting temperature & $\unit{1809}{\kelvin}$\\
			$H_f$ & Heat of fusion & $\unit{2.471\cdot10^{5}}{\joule\per\kilo\gram}$ \\
			$\rho_\mathrm{liq}/\rho_\mathrm{solid}$ & Ratio of liquid to solid density at melt point & $0.955$ \\
			\hline
		\end{tabular}
		
		\caption{Material parameters for iron with ANEOS equation of state. Other parameters are set to zero.}
		\label{tab:aneos-iron}
	\end{center}
\end{table}

For a more physical computation of temperature, and to handle phase transitions in a more thermodynamically consistent way than is done with Tillotson EoS, we use ANEOS \citep{ANEOS} for iron with the parameters provided in table~\ref{tab:aneos-iron}, and M-ANEOS \citep{2007M&PSMelosh} for silicates. These equations of state provide numerous output variables including the temperature and phase information (e.g. melt and vapor fraction). Depending on the parameters used in the equation of state, this phase information can be given in different ways. For iron, it can be used to infer the melting temperature at a given pressure. However, this is not the case for silicates, for which we apply the same procedure as used in \citet{2009E&PSLSenftStewart} to obtain the melting temperature at a given pressure.

It is important to point out that neither the Tillotson nor the ANEOS equation of state include the latent heat of melting and therefore, temperatures above the melt temperature are overestimated.

Shocks occur when the impact speed (or particle velocity) exceeds the sound speed in the medium. If not treated properly, this can lead to particle interpenetration and other non-physical effects. We use a standard artificial viscosity term \citep{1992ARA&AMonaghan} that deposits internal energy and momentum behind the shock in a manner that is consistent with the Hugoniot shock relations \citep[c.f.][]{1989BookMelosh} but spread out over several smoothing lengths.

In addition to the body force accelerations given by equation (\ref{eq:model:velmat}), we also compute the accelerations due to self-gravity, using a tree-code method \citep{1986NatureBarnesHut}.

The set of equations described above provides the time dependence of the physical quantities, which are then integrated using a prediction-correction scheme \citep{NumRecCPP}. For a more detailed description of the method see e.g. \citet{2015BookJutzi}.

% --------------------------
\section{Initial conditions}
\label{sec:initialconds}
% --------------------------
\subsection{Setup}
We perform a series of SPH collision models, where we use a Mars-mass target body ($r=\unit{3400}{\kilo\meter}$) and an impactor with $\unit{1000}{\kilo\meter}$ radius. 
We use different EoS (ANEOS or Tillotson), material rheologies (solid or fluid) and numerical schemes (density summation or integration), in order to study the effects on the collision outcome and to asses their relative importance. For each combination, we use two impact angles: 0\textdegree{} (head-on) or 45\textdegree{} (oblique), which gives a total of 16 simulations. The mutual escape velocity \citep{2010ChEGAsphaug} defines the impact speed. As a nominal case, we define the simulation with ANEOS, solid rheology, integrated density and oblique impact angle. 

The target and impactor both start with a Mars-like internal structure with an iron core radius half of the body radius, while SiO2 comprises the bodies’ mantles. The use of SiO2 to represent Mars’ basaltic mantle is a simplified but reasonable assumption, given available equation of state information. Note also that the crust of the target is not resolved at these large scales and is not explicitly included in the model.
For simulations using Tillotson EoS, each layer is initially isothermal; iron core temperature is fixed to $T_\mathrm{Fe} = \unit{1800}{\kelvin}$ and mantle to $T_\mathrm{Si} = \unit{1500}{\kelvin}$. For the simulations with ANEOS, an isentropic profile is used for which core's central temperature is $\unit{1800}{\kelvin}$ and for the silicate at the core-mantle boundary it is $\unit{1500}{\kelvin}$. The choice of these values is explained in \citet{Companion}.

We begin by setting up self-gravitating, hydrostatically-equilibrated planets, starting with one-dimensional (1D) spherically symmetric bodies modelled using a  Lagrangian hydrocode \citep{1989BookBenz} with the same EoS as in the SPH model. This structure is divided in \numprint{1000} cells (100 for the impactor) of equal mass. Cell boundaries according to the force balance between self-gravity and pressure (including a damping term).
This profile is evolved until hydrostatic equilibrium is reached so that radial velocities are small (less than 1\% of the escape velocity). Afterwards we transfer the 1D radial profile onto SPH particles that are placed onto a 3D lattice. Parameters of each particle are copied from the profile according to the radius. As particles are equally spaced on the lattice, variation of density is taken into account by adjusting the particle mass. The spherical SPH bodies are then also evolved in an initialising step to reach a hydrostatic equilibrium and negligible radial velocities. Thus the SPH simulations start with two relaxed differentiated spherical planets approaching from several radii away;  they get tidally deformed prior to the impact.

The SPH simulations are performed with a resolution of about one million particles for the target. The number of particles for the impactor is scaled according to the mass ratio between the two objects, so that particle spacing $h$ is approximately constant, for the greatest numerical accuracy during the collision. The corresponding smoothing length is then approximately $\unit{60}{\kilo\meter}$ for both bodies.

\subsection{Initial radial profiles}
\begin{figure}
\makebox[\textwidth][c]{
\includegraphics{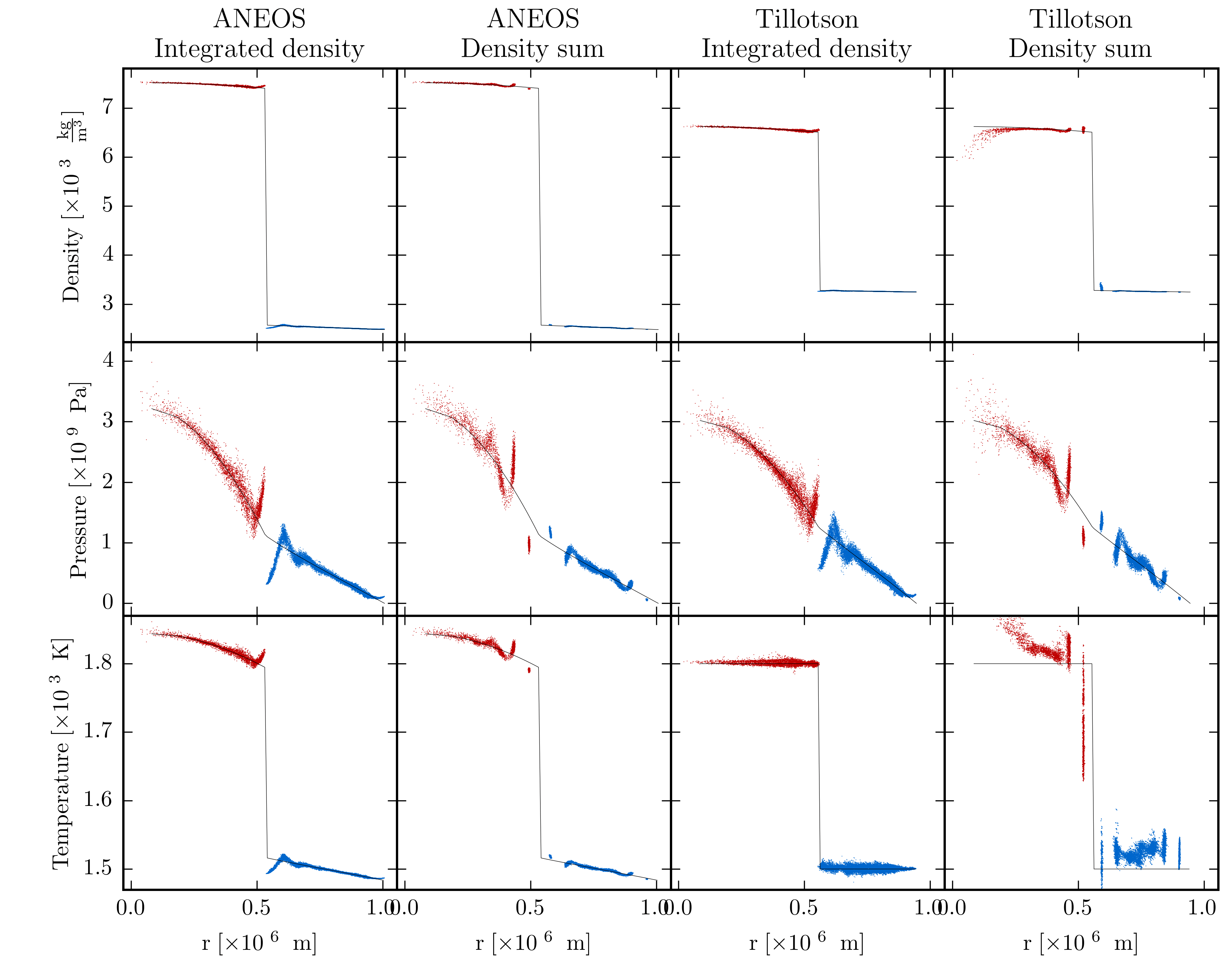}
}
\begin{center}
\caption{Initial profiles of the different impactors used in this study. The 1D profile is depicted by the black line. Each point is one SPH particle and its color represents the material: red for iron and blue for silicate.}
\label{fig:profile:impactor}
\end{center}
\end{figure}

\begin{figure}
\makebox[\textwidth][c]{
\includegraphics{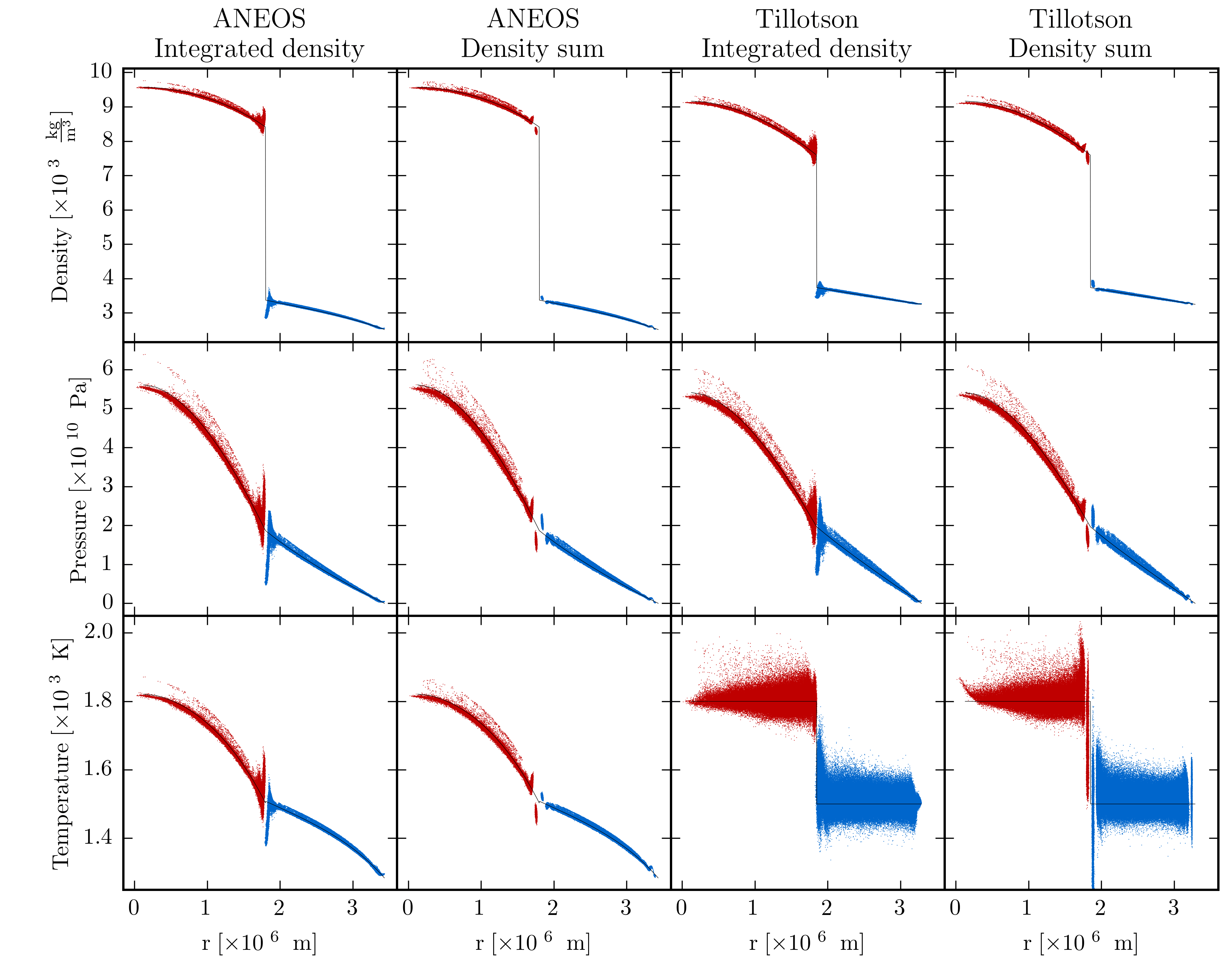}
}
\begin{center}
\caption{Initial profiles of the different targets used in this study.}
\label{fig:profile:target}
\end{center}
\end{figure}

Initial radial profiles of density, pressure and temperature are shown in figures~\ref{fig:profile:impactor} and~\ref{fig:profile:target} for the impactor and target respectively. For the cases with the Tillotson EoS, the internal energy is being evolved during the equilibration  phase, hence the spread in temperature. The comparison with a case where this value is kept fixed is discussed later in section~\ref{sec:tillotson-profile}. The 1D profiles for the two different approaches to compute the density (density sum and density integration) are identical as the distinction only happens during the SPH phase itself.

There is a noticeable difference in the density between the cases with a different EoS: iron has higher density with ANEOS and the opposite happens for silicate. The density contrast at the core-mantle boundary is thus higher when using ANEOS. Since the objects have the same mass in the two cases, their radius is affected by this density variation and we obtain about $\unit{3440}{\kilo\meter}$ using ANEOS and $\unit{3290}{\kilo\meter}$ using Tillotson, which makes a relative difference of the radii around $\unit{4}{\%}$.
It is known that standard SPH has problems handling sharp density changes, at boundaries between materials and at the surface. We see artificial numerical effects at these locations. With density summation, particles beyond the boundary enter in the SPH sum (equation \ref{eq:sphsum}) and thus smooth the discontinuity.  Particles in this region then have a slightly different density. At the surface, the effect is due to the lack of close-by particles (see \citealp{2017MNRASReinhardt} for an improved scheme). The density variation is reflected in pressure and temperature.
Although in the case of density integration there is no such summation involved to compute the density, there are still some artificial effects (oscillations) at the boundaries. More sophisticated ways to deal with discontinuous boundaries in SPH have been developed recently \citep[e.g.][]{2013PASJHosonoSaitohMakino}  and shall be implemented in our code for future studies.

However, as we shall see below, the amplitude of the unphysical 'noise' in the initial profiles is small enough not to affect significantly the outcome of the collision simulations. 

% ---------------
\section{Results}
% ---------------

\subsection{Nominal case}
\label{sec:results-nomcase}

\begin{figure}
\def\arraystretch{0.}
\makebox[\textwidth][c]{
\begin{tabularx}{6.5in}{@{}cc@{}c@{}c@{}c@{}}
 & Solid, material & Solid, temperature & Fluid, material & Fluid, temperature \\
\rotatebox[origin=c]{90}{T=$\unit{0.00}{\hour}$} & \multicolumn{1}{@{}m{1.5in}@{}}{\includegraphics{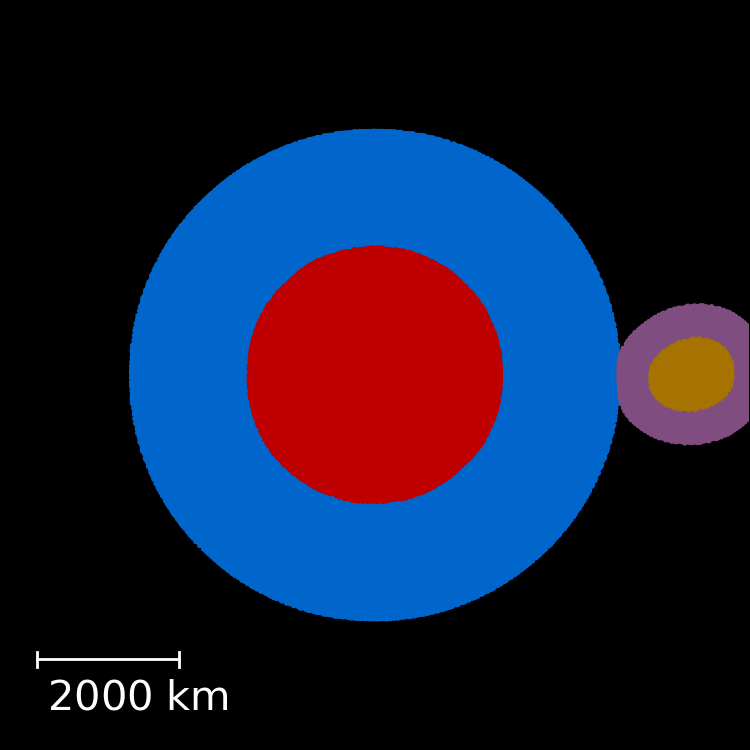}} & \multicolumn{1}{@{}m{1.5in}@{}}{\includegraphics{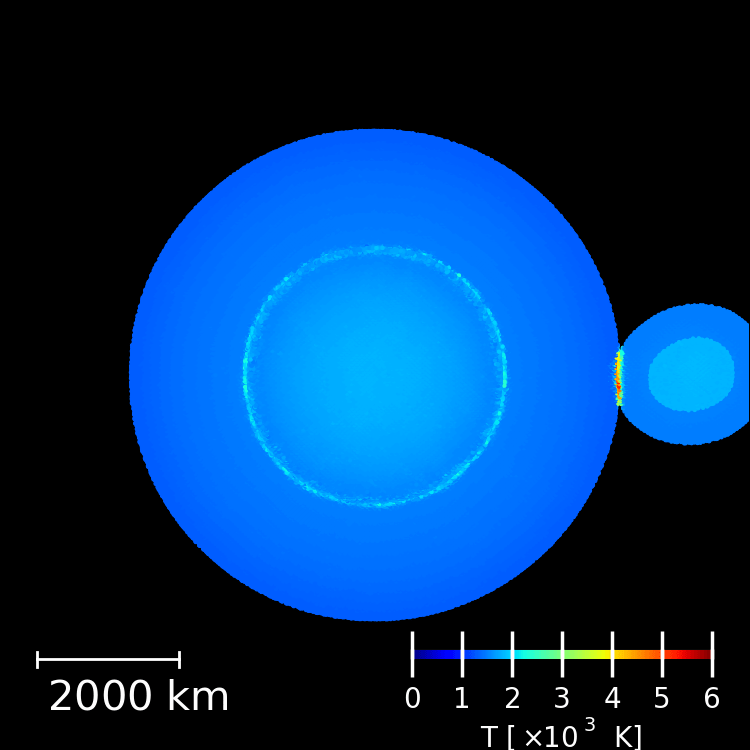}} & \multicolumn{1}{@{}m{1.5in}@{}}{\includegraphics{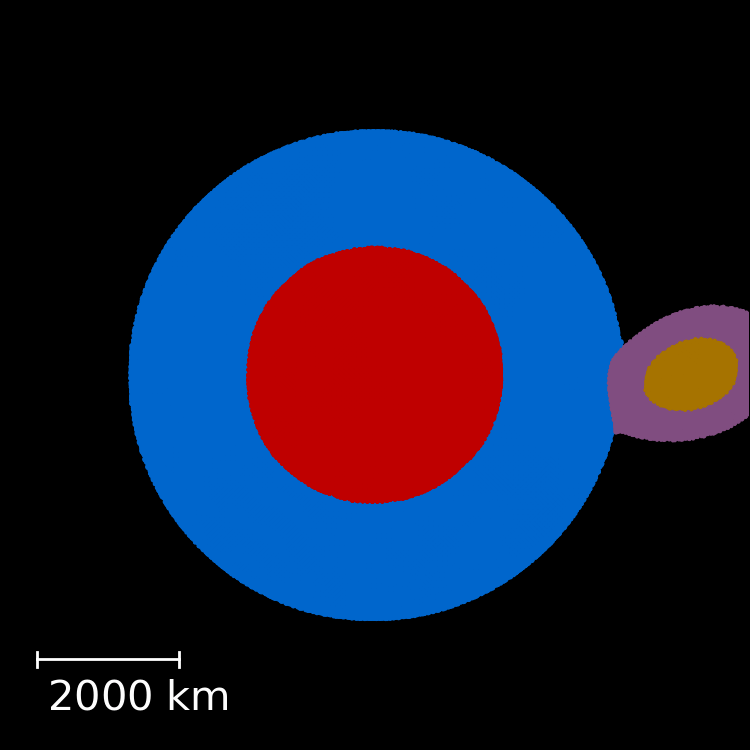}} & \multicolumn{1}{@{}m{1.5in}@{}}{\includegraphics{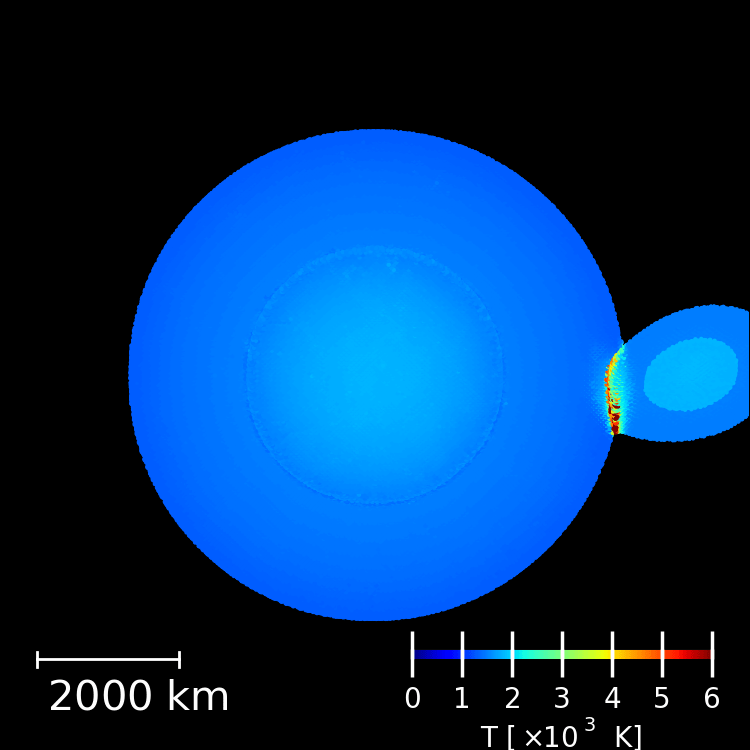}} \\
\rotatebox[origin=c]{90}{T=$\unit{0.25}{\hour}$} & \multicolumn{1}{@{}m{1.5in}@{}}{\includegraphics{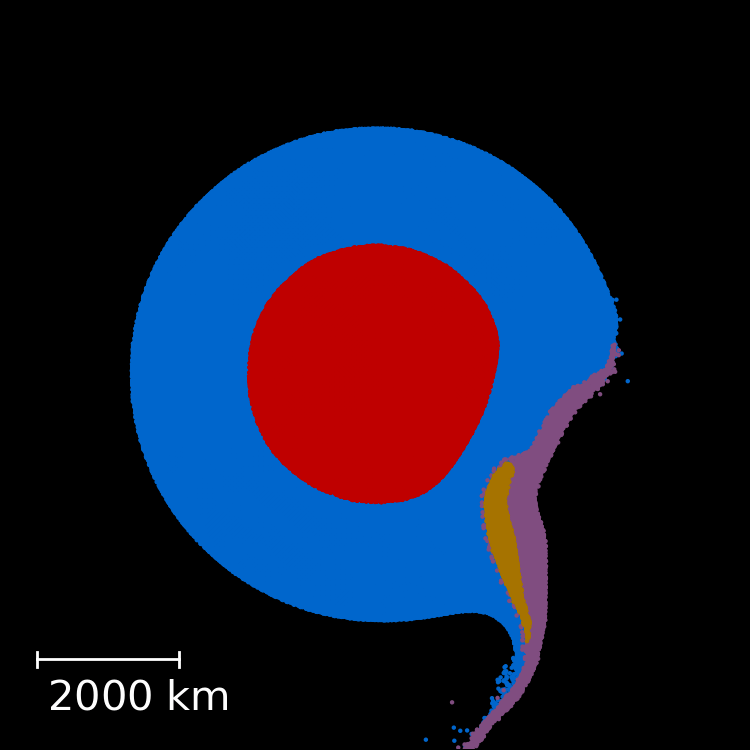}} & \multicolumn{1}{@{}m{1.5in}@{}}{\includegraphics{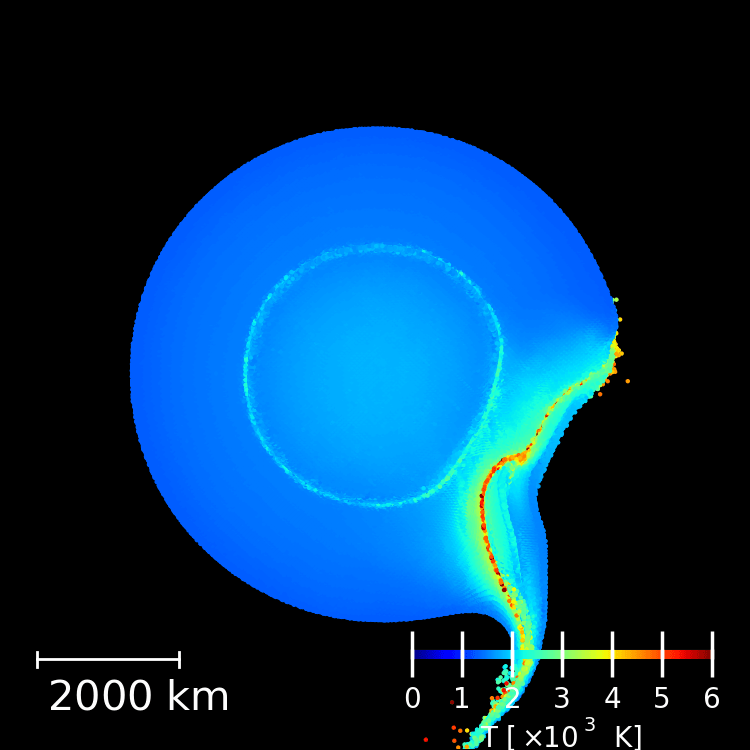}} & \multicolumn{1}{@{}m{1.5in}@{}}{\includegraphics{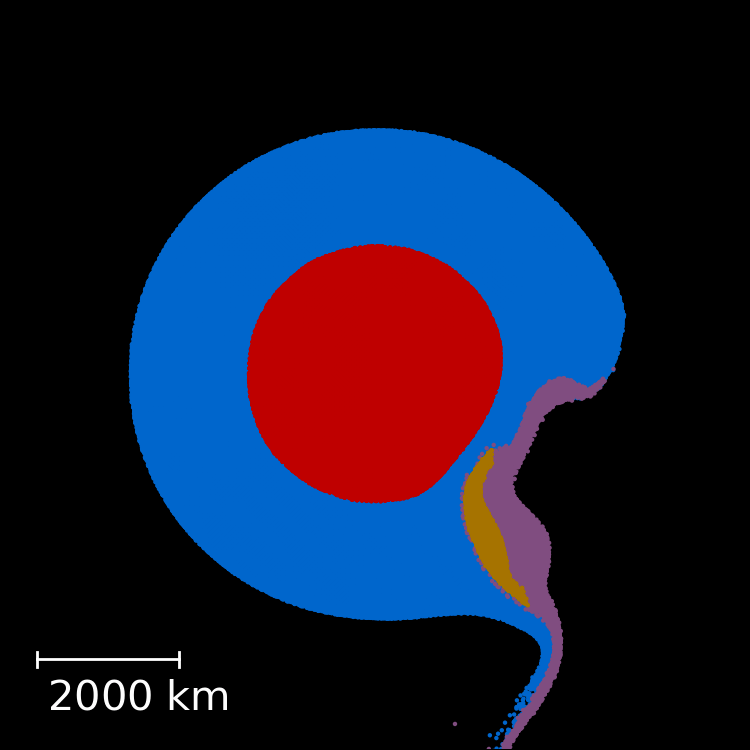}} & \multicolumn{1}{@{}m{1.5in}@{}}{\includegraphics{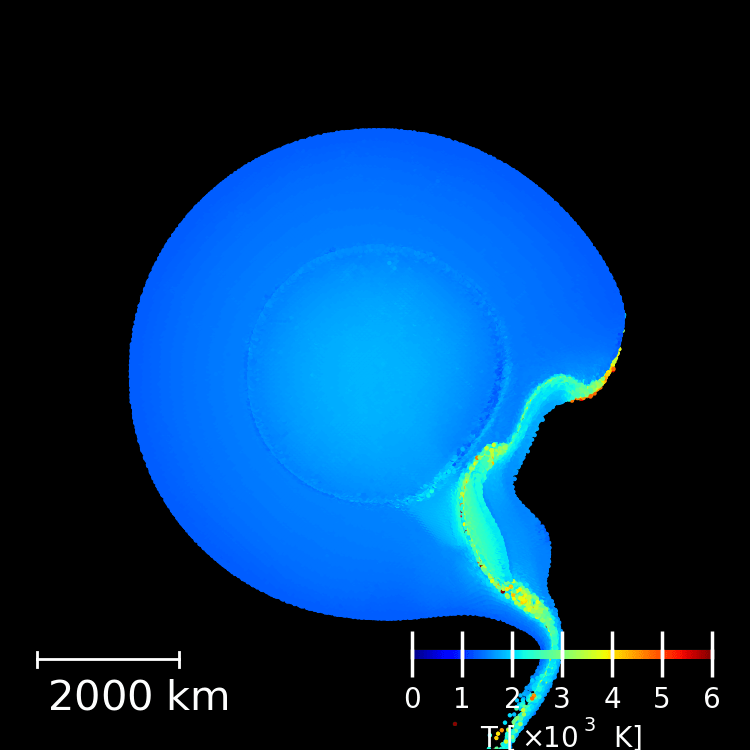}} \\
\rotatebox[origin=c]{90}{T=$\unit{0.50}{\hour}$} & \multicolumn{1}{@{}m{1.5in}@{}}{\includegraphics{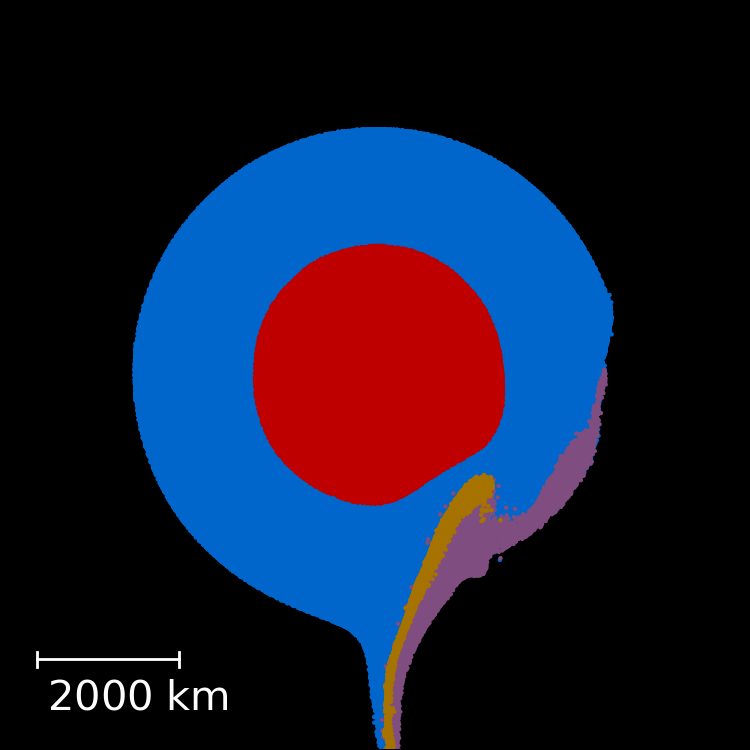}} & \multicolumn{1}{@{}m{1.5in}@{}}{\includegraphics{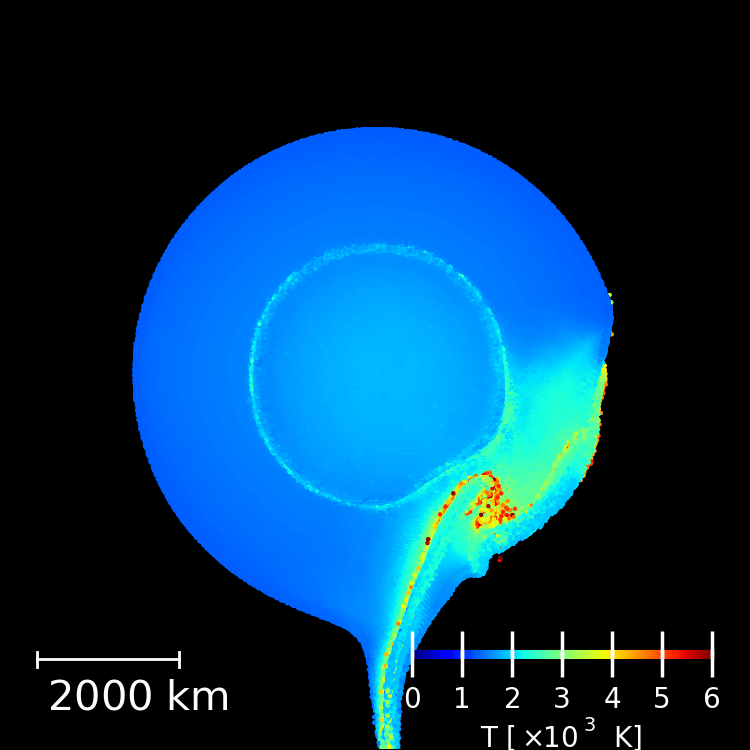}} & \multicolumn{1}{@{}m{1.5in}@{}}{\includegraphics{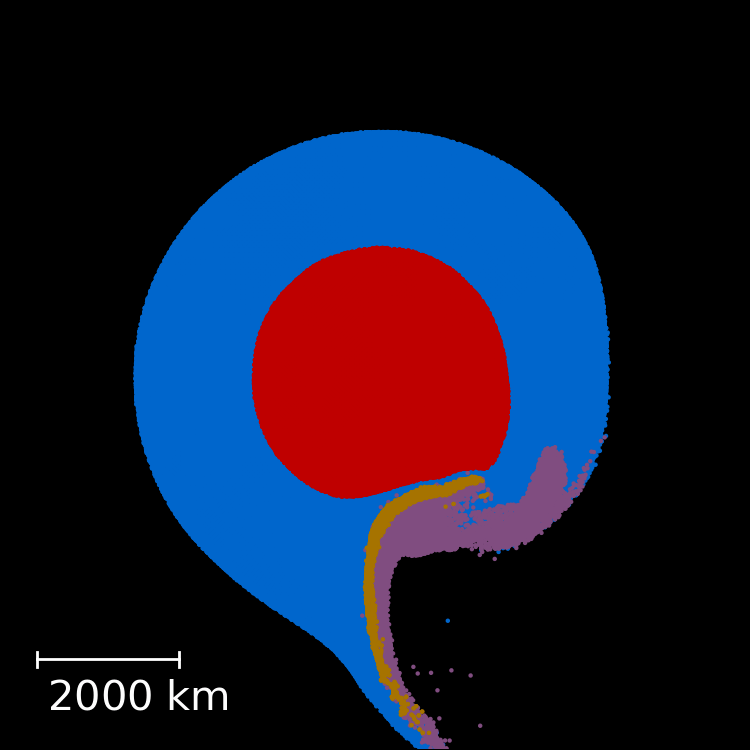}} & \multicolumn{1}{@{}m{1.5in}@{}}{\includegraphics{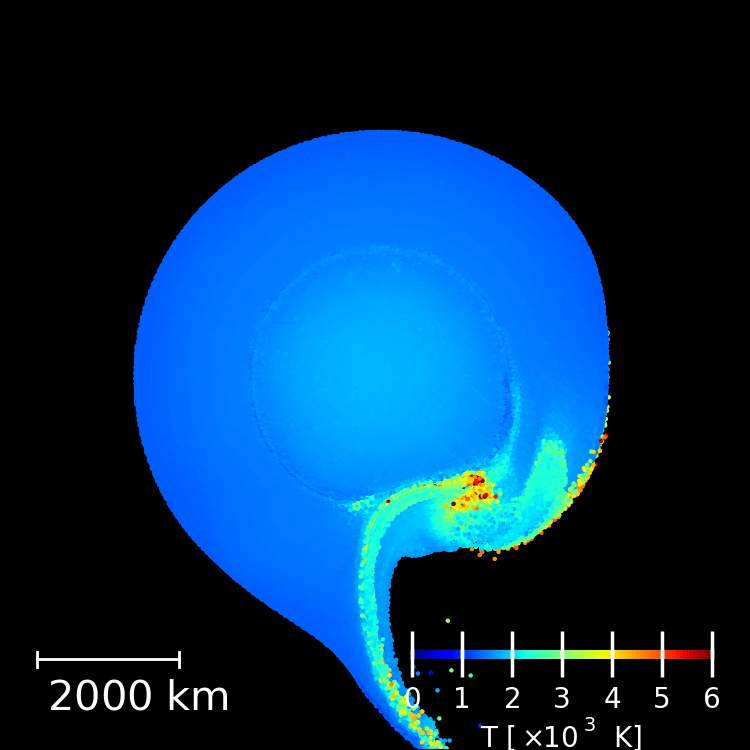}} \\
\rotatebox[origin=c]{90}{T=$\unit{0.92}{\hour}$} & \multicolumn{1}{@{}m{1.5in}@{}}{\includegraphics{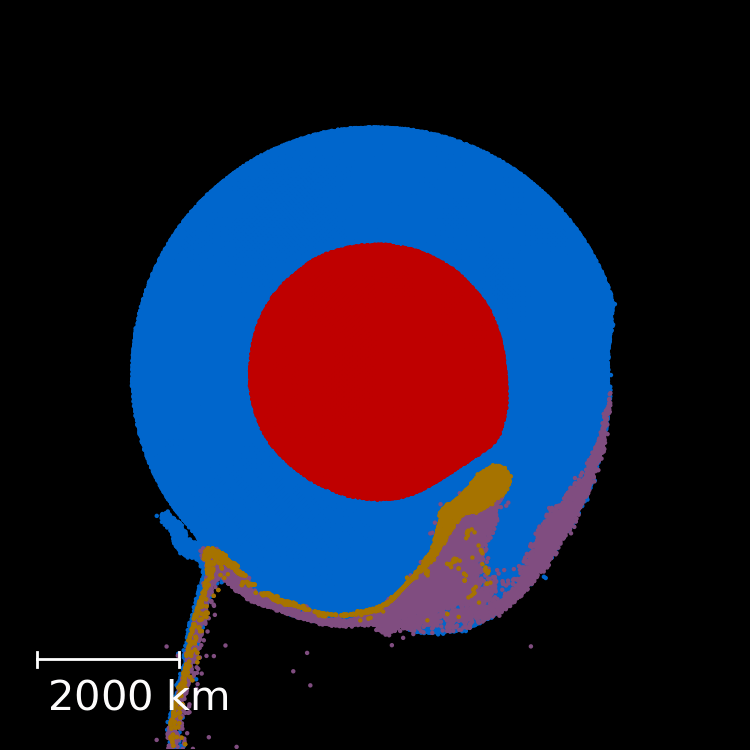}} & \multicolumn{1}{@{}m{1.5in}@{}}{\includegraphics{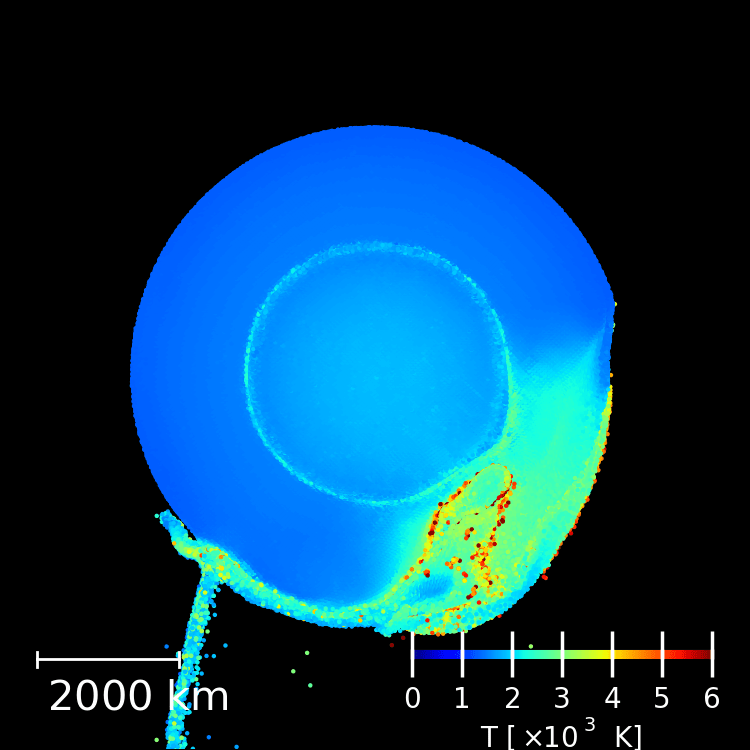}} & \multicolumn{1}{@{}m{1.5in}@{}}{\includegraphics{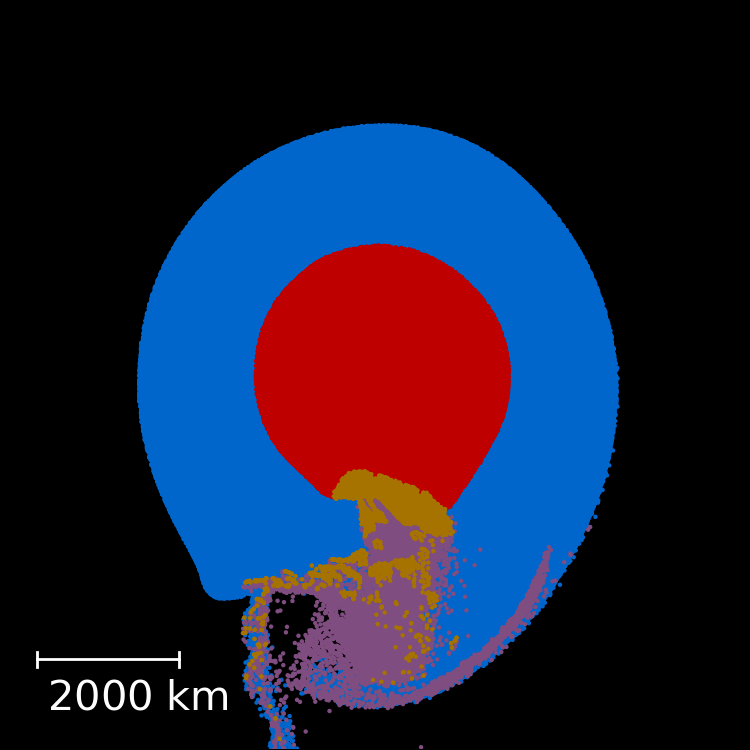}} & \multicolumn{1}{@{}m{1.5in}@{}}{\includegraphics{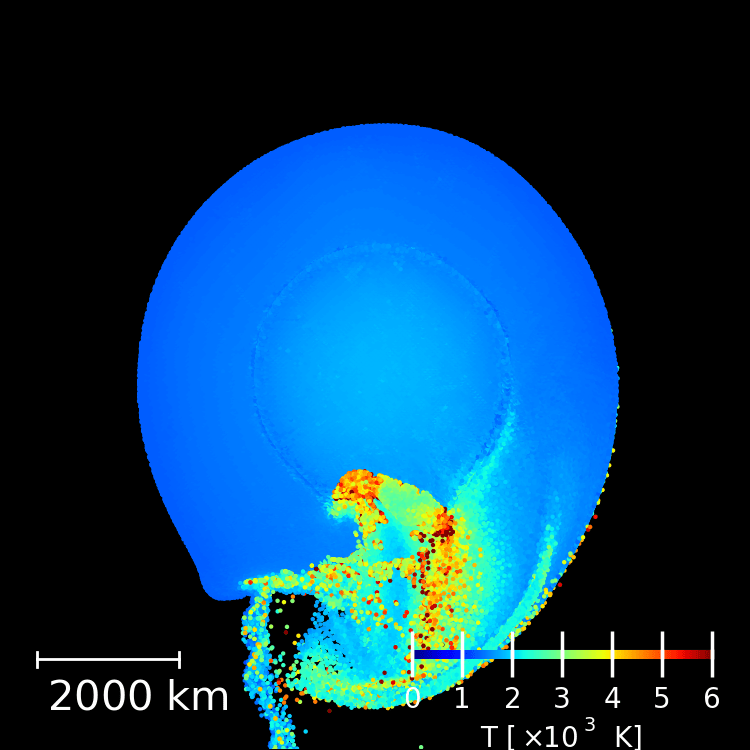}} \\
\rotatebox[origin=c]{90}{T=$\unit{1.33}{\hour}$} & \multicolumn{1}{@{}m{1.5in}@{}}{\includegraphics{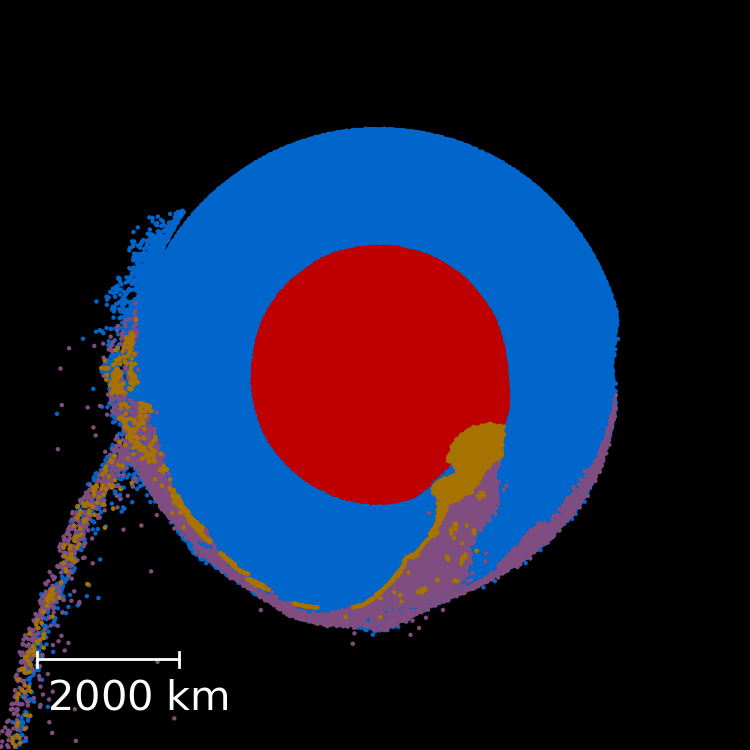}} & \multicolumn{1}{@{}m{1.5in}@{}}{\includegraphics{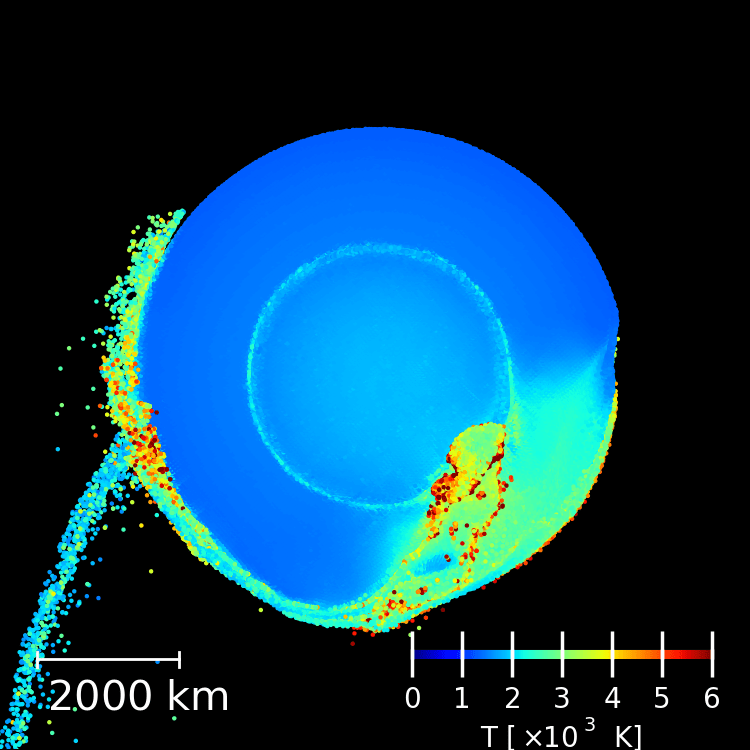}} & \multicolumn{1}{@{}m{1.5in}@{}}{\includegraphics{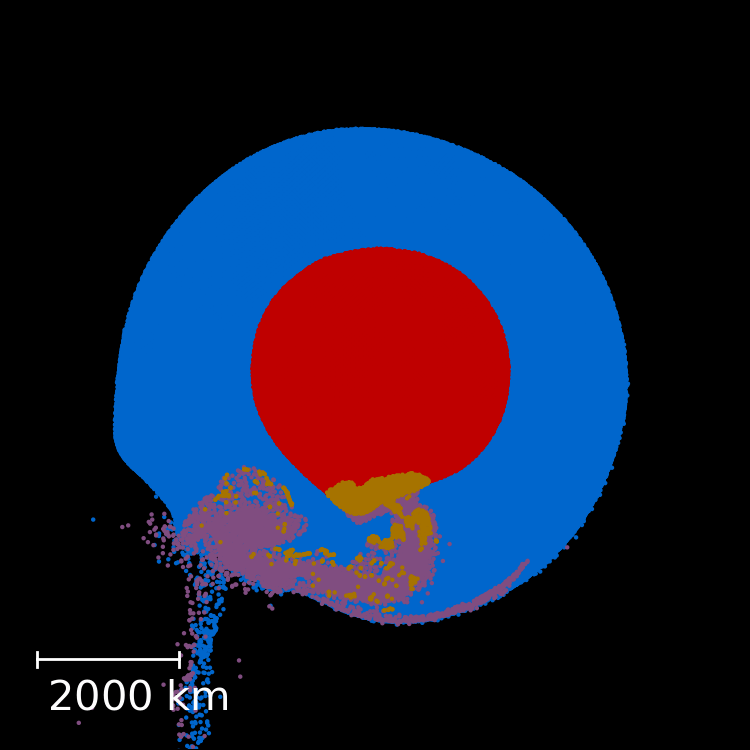}} & \multicolumn{1}{@{}m{1.5in}@{}}{\includegraphics{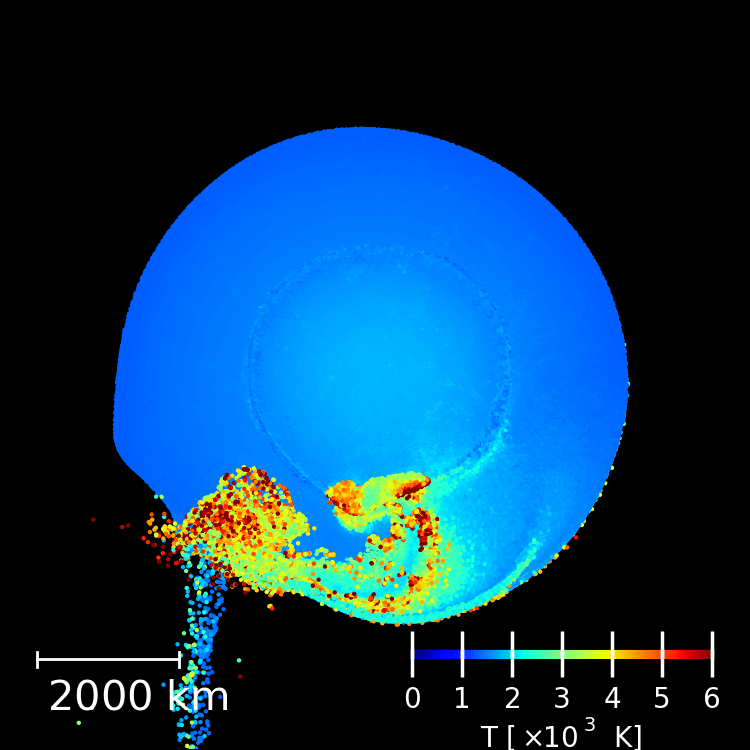}} \\
\end{tabularx}
}
%\begin{center}
%\end{center}
\end{figure}

\begin{figure}
\def\arraystretch{0.}
\makebox[\textwidth][c]{
\begin{tabularx}{6.5in}{@{}cc@{}c@{}c@{}c@{}}
 & Solid, material & Solid, temperature & Fluid, material & Fluid, temperature \\
\rotatebox[origin=c]{90}{T=$\unit{3.00}{\hour}$} & \multicolumn{1}{@{}m{1.5in}@{}}{\includegraphics{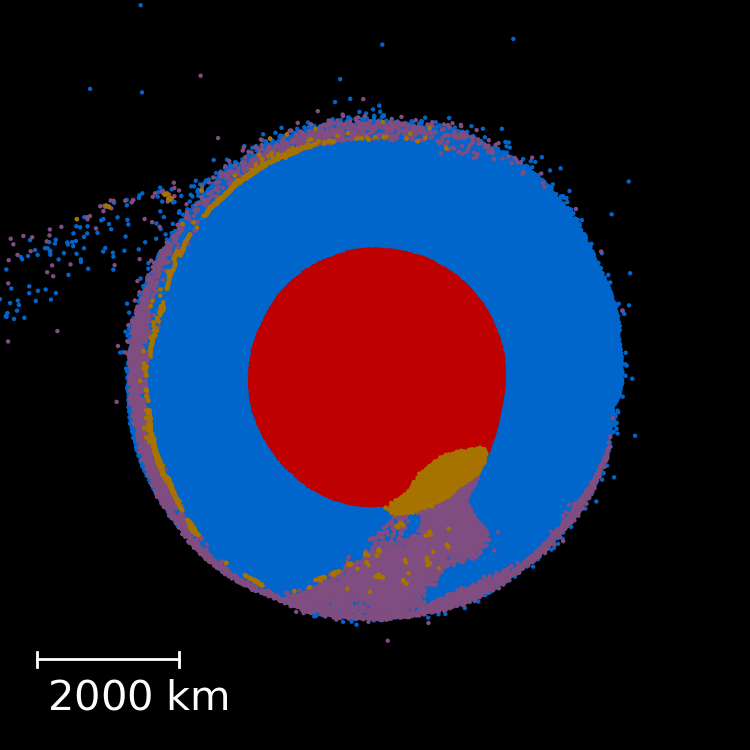}} & \multicolumn{1}{@{}m{1.5in}@{}}{\includegraphics{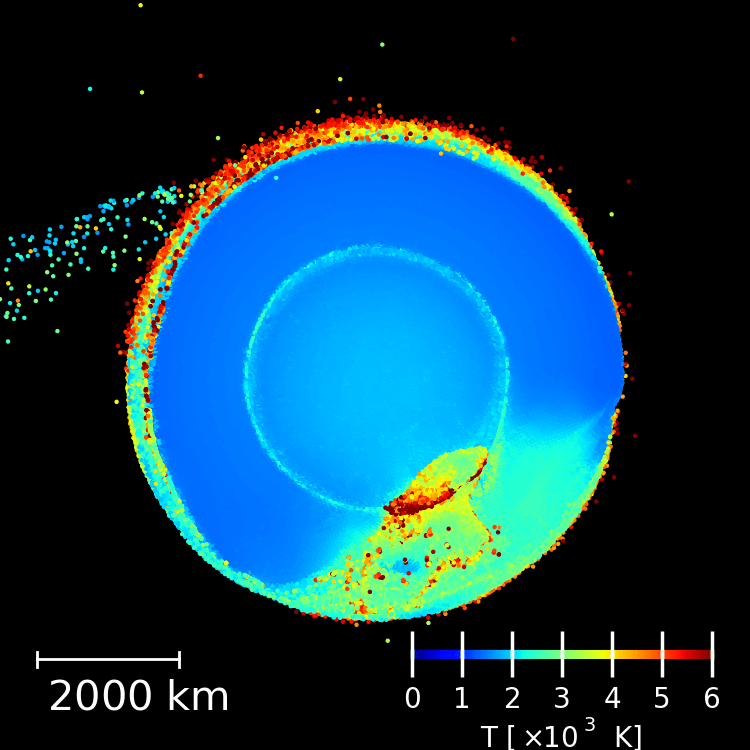}} & \multicolumn{1}{@{}m{1.5in}@{}}{\includegraphics{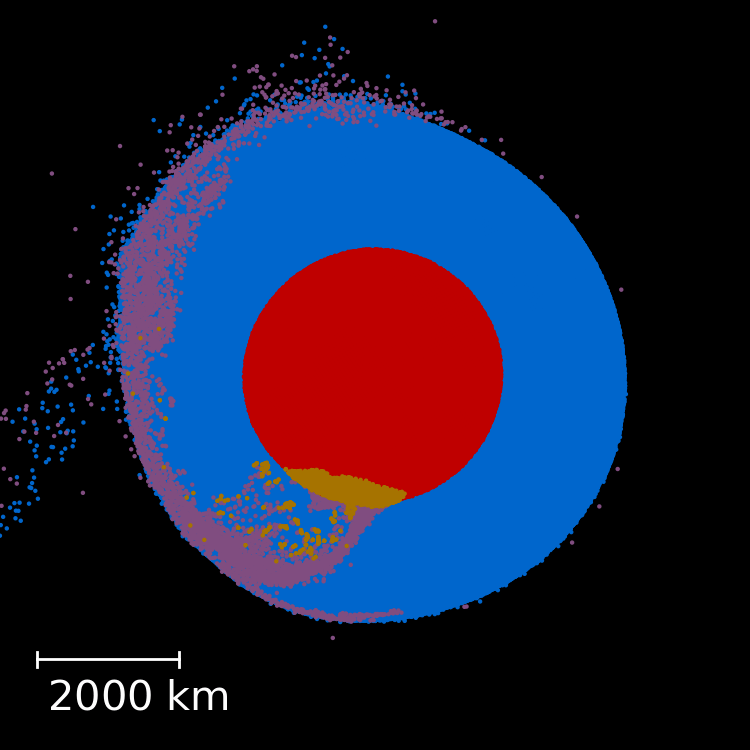}} & \multicolumn{1}{@{}m{1.5in}@{}}{\includegraphics{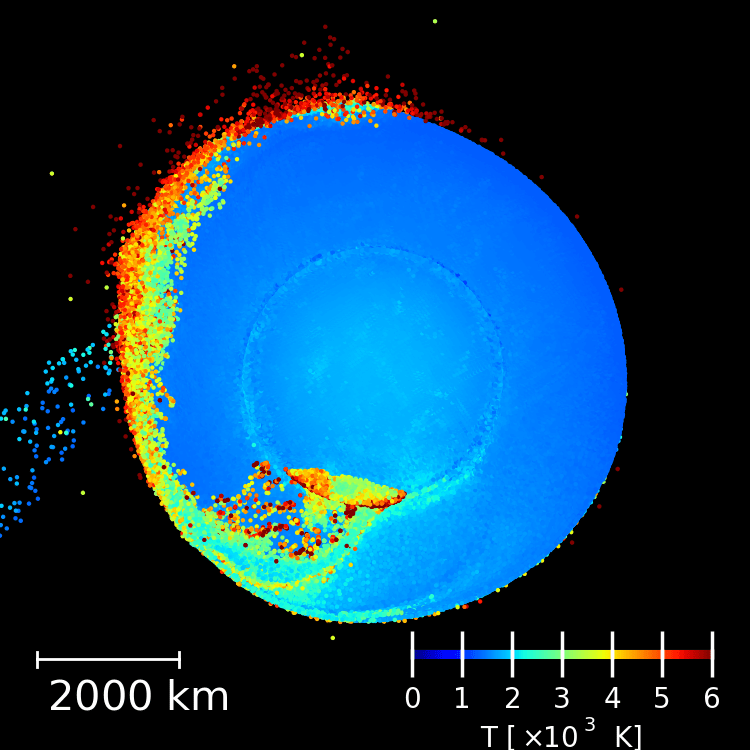}} \\
\rotatebox[origin=c]{90}{T=$\unit{4.00}{\hour}$} & \multicolumn{1}{@{}m{1.5in}@{}}{\includegraphics{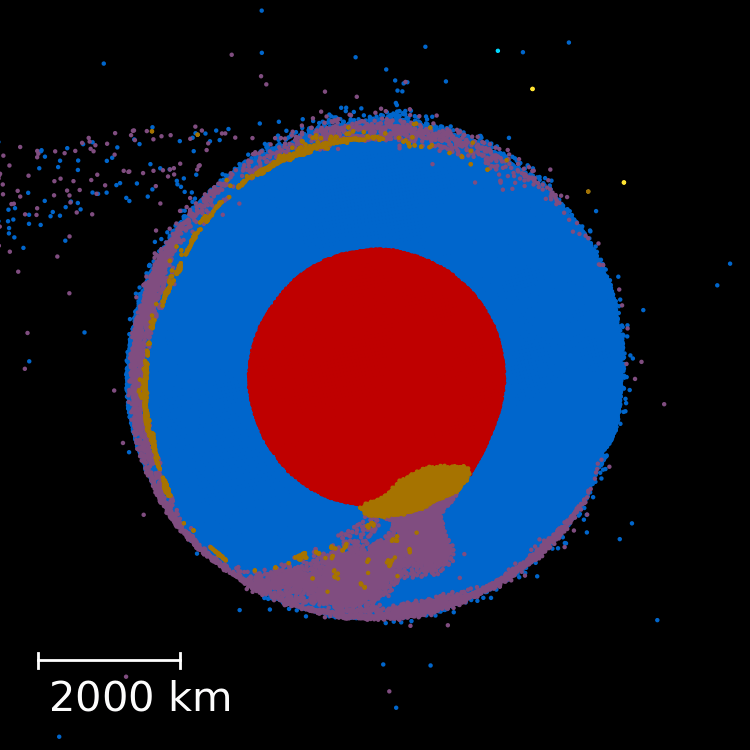}} & \multicolumn{1}{@{}m{1.5in}@{}}{\includegraphics{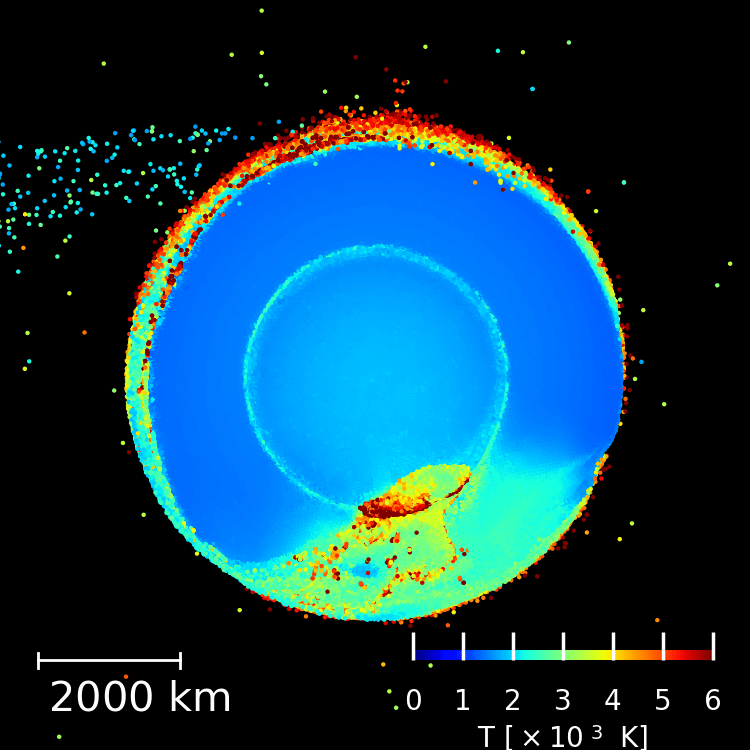}} & \multicolumn{1}{@{}m{1.5in}@{}}{\includegraphics{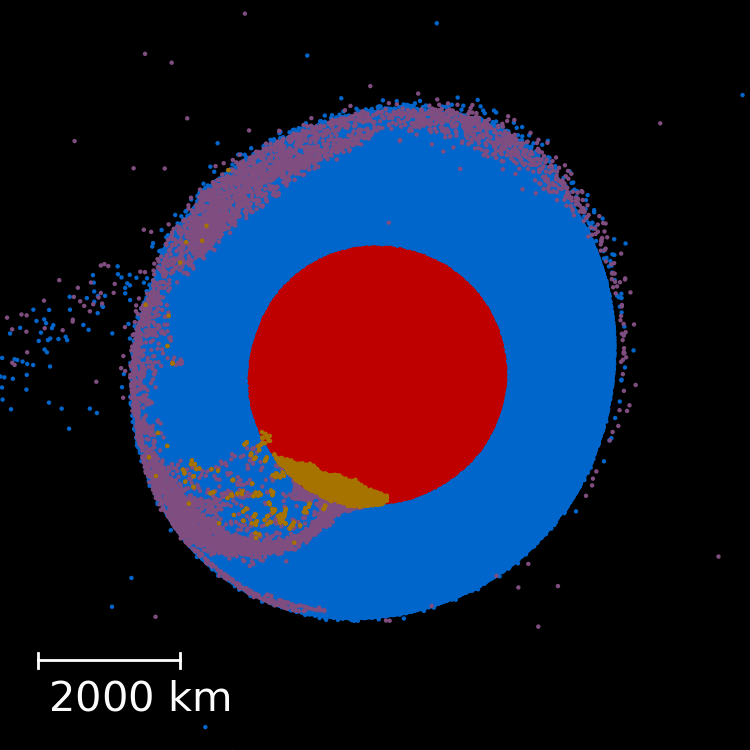}} & \multicolumn{1}{@{}m{1.5in}@{}}{\includegraphics{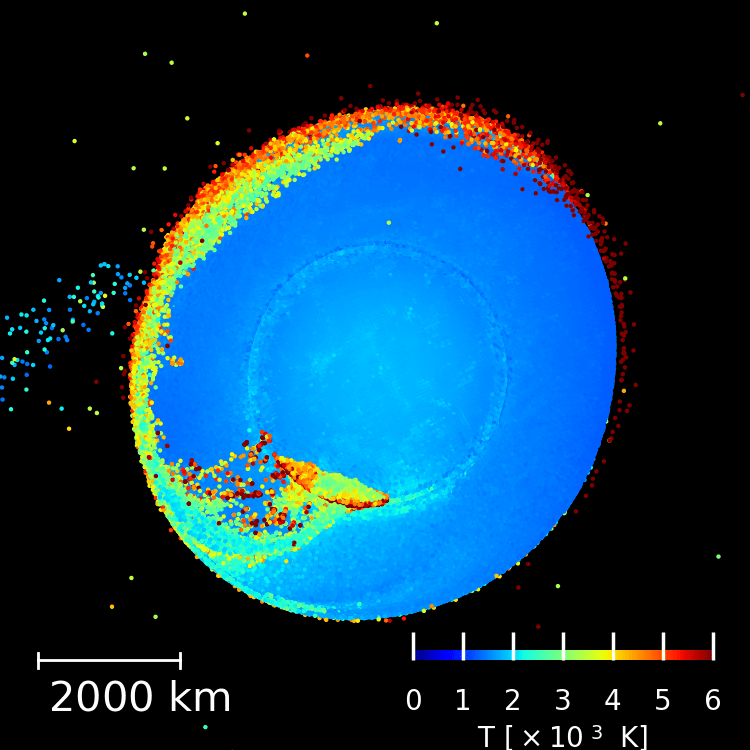}} \\
\rotatebox[origin=c]{90}{T=$\unit{18.00}{\hour}$} & \multicolumn{1}{@{}m{1.5in}@{}}{\includegraphics{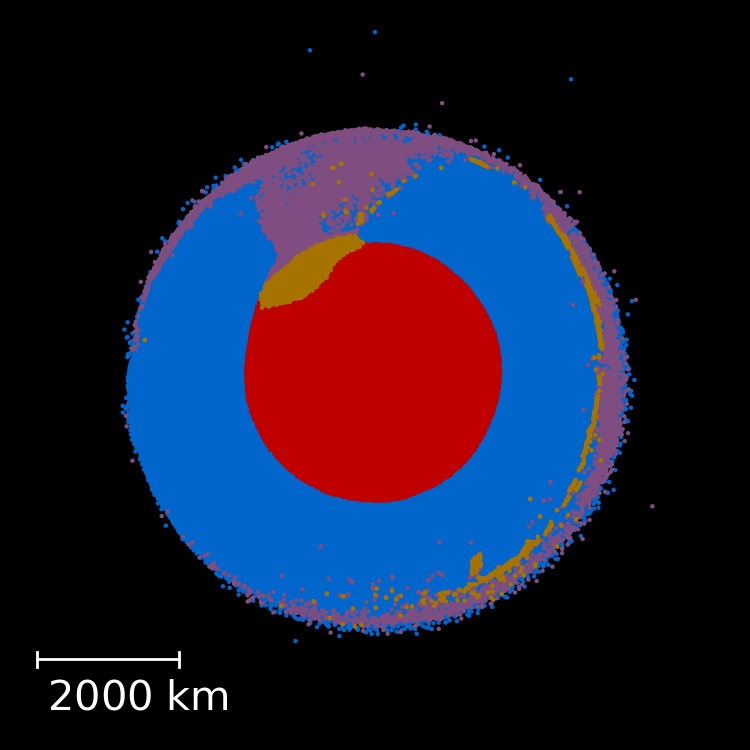}} & \multicolumn{1}{@{}m{1.5in}@{}}{\includegraphics{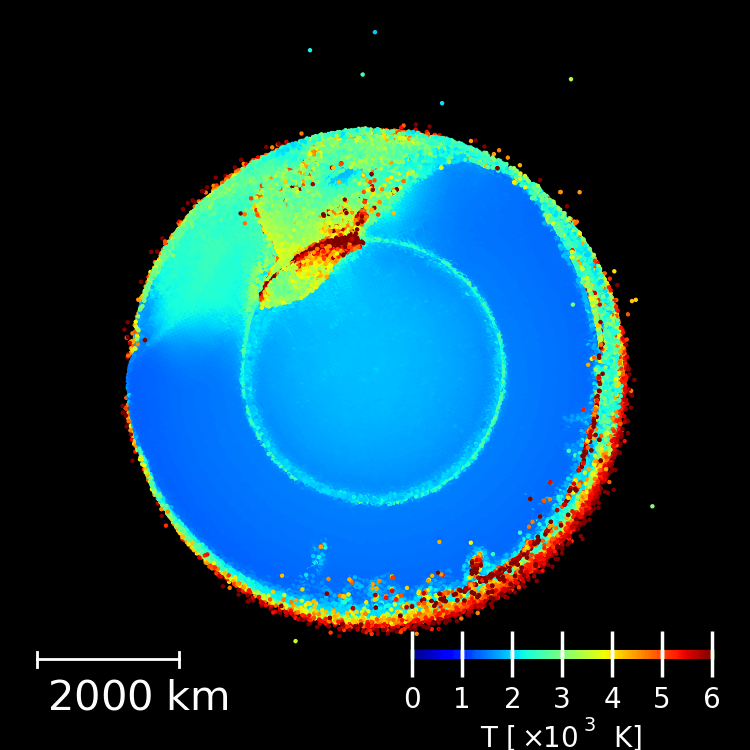}} & \multicolumn{1}{@{}m{1.5in}@{}}{\includegraphics{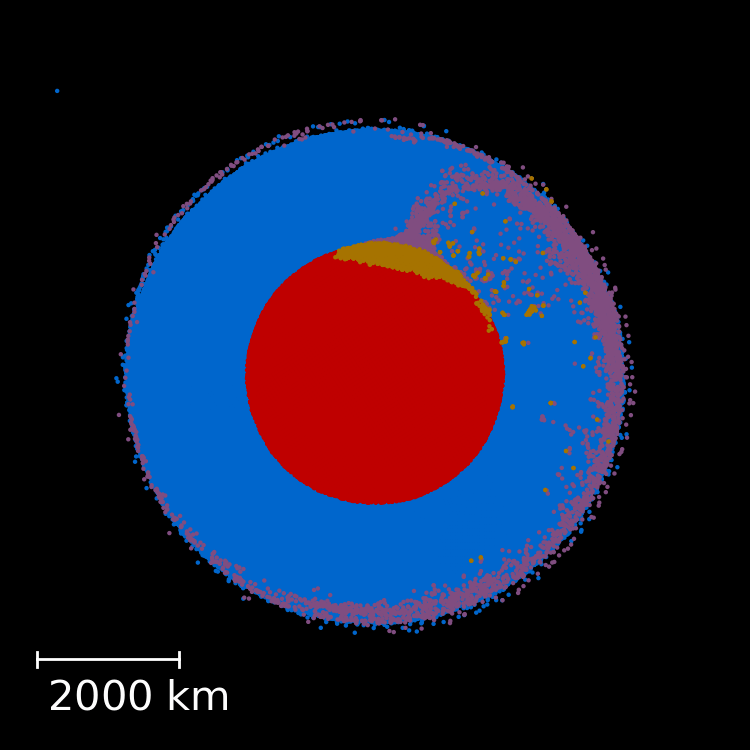}} & \multicolumn{1}{@{}m{1.5in}@{}}{\includegraphics{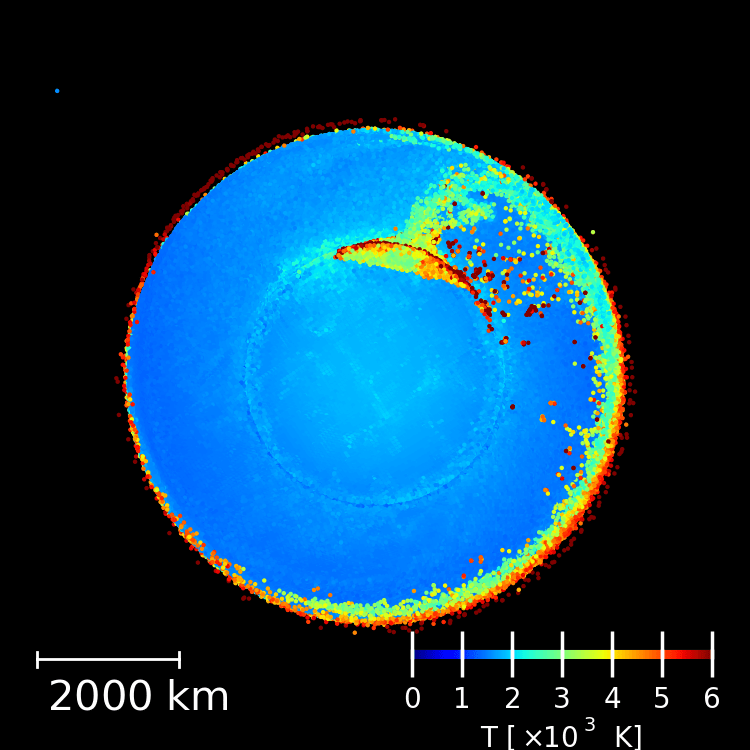}} \\
\end{tabularx}
}
\begin{center}
\caption{Time series showing material distribution for the nominal case ($\unit{45}{\degree}$, ANEOS, solid, integrated density; two left columns) and the corresponding fluid case (two right columns). Impactor's trajectory is clockwise. This is a slice of $\unit{1000}{\kilo\meter}$ depth inside the impact plane. Plots are centered on the center of mass of the main body. On material plots, colour represents the type and origin: blue is target's mantle, purple impactor's mantle, red target's core and yellow impactor's core.}
\label{fig:nomcase:timeseries}
\end{center}
\end{figure}

We show a time series of the nominal case along with the corresponding fluid case in figure~\ref{fig:nomcase:timeseries}. Shown are the post-impact material and temperature distributions. The simulations start roughly two hours before the first snapshot. Due to the angular momentum transferred to the target by the impact, the target begins to rotate after a few hours, hence the impact location is shifted by $\unit{\sim180}{\degree}$ between the two last snapshots. One can note the impactor's tidal deformation before the impact mainly in the fluid case. The effect of the material rheology leads to significant differences in terms of the post-impact material and temperature distribution. Another notable difference is the degree of the impact induced target oscillation, which is much stronger in the fluid case (see snapshots at 3 and $\unit{4}{\hour}$).

The heat generated by the impact is spread out over a broader region with solid rheology compared to the fluid one. Already at $t=\unit{0.25}{\hour}$ the higher temperature zone extend to several hundreds of kilometres away from the contact zone and expands until covering a wide region at the end of our simulations. With fluid rheology only the contact zone has high temperature material at the beginning and towards the end only impactor material retains these high temperatures. Note that the annulus of warm material ($\unit{\sim3000}{\kelvin}$) that forms at the target's core-mantle boundary with solid rheology is not due to the impact but is an artificial feature caused by the set up routine. The small remaining velocities (on the order of $\unit{10}{\meter\per\second}$) at the beginning of the impact simulation,  when forces are enabled and the damping term removed are sufficient to create instabilities that lead to internal energy increase due to the artificial viscosity.

At early times, the impactor does not penetrate as deep inside the target with solid rheology  as in the fluid case (see snapshots at $\unit{0.50}{\hour}$ and  $\unit{0.92}{\hour}$).

\begin{figure}
	\begin{center}
		\includegraphics{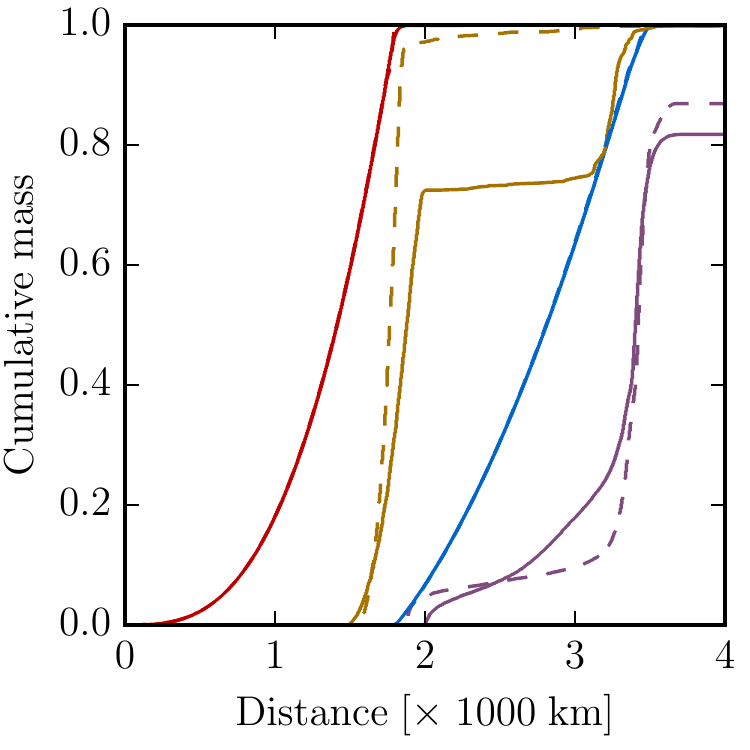}
		\caption{Cumulative mass distribution at the end of the simulation as function of radius. Colour represents material type and origin: red is target's core, yellow impactor's core, blue target's mantle and purple impactor's mantle, as for material plots in figure~\ref{fig:nomcase:timeseries}. Solid lines depict the run with solid rheology and dashed lines are for fluid rheology.}
		\label{fig:cumulmass:canonical}
	\end{center}
\end{figure}

To get a better overview of the location where impactor material is deposited inside the target at the end of the simulation, we show a cumulative mass plot against radius in figure~\ref{fig:cumulmass:canonical}. The curves are normalised by the total mass of each material.

Note that some of the impactor's silicate material is still orbiting around the body or has been ejected.  This accounts for $\unit{13}{wt\%}$ and $\unit{18}{wt\%}$ for the fluid and solid rheologies respectively. In both cases, about one sixth of this material is still bound whereas the remaining is on an escaping orbit.

There are significant differences for impactor's material distribution between the two rheologies. $\unit{27}{wt\%}$ of impactor's core remains inside the mantle with solid rheology whereas this value falls to  $\unit{3}{wt\%}$ for fluid rheology. The remaining lies at the target's core-mantle boundary. The impactor's silicate shows a different behaviour. For fluid rheology, there is a small part close to the core-mantle boundary, some inside the target's mantle and the majority close to the surface. For solid rheology, there is one half distributed inside the mantle and the other half close to the surface.

\subsection{Comparison of EoS and material rheologies}

\begin{figure}
\def\arraystretch{0.}
\makebox[\textwidth][c]{
\begin{tabularx}{6.5in}{@{}cc@{}c@{}c@{}c@{}}
 & Solid, material & Solid, temperature & Fluid, material & Fluid, temperature \\
\rotatebox[origin=c]{90}{\parbox{1.5in}{ANEOS\\density integration}} & \multicolumn{1}{@{}m{1.5in}@{}}{\includegraphics{ao_64800_material}} & \multicolumn{1}{@{}m{1.5in}@{}}{\includegraphics{ao_64800_temperature}} & \multicolumn{1}{@{}m{1.5in}@{}}{\includegraphics{fo_64800_material}} & \multicolumn{1}{@{}m{1.5in}@{}}{\includegraphics{fo_64800_temperature}} \\
\rotatebox[origin=c]{90}{\parbox{1.5in}{ANEOS\\density summation}} & \multicolumn{1}{@{}m{1.5in}@{}}{\includegraphics{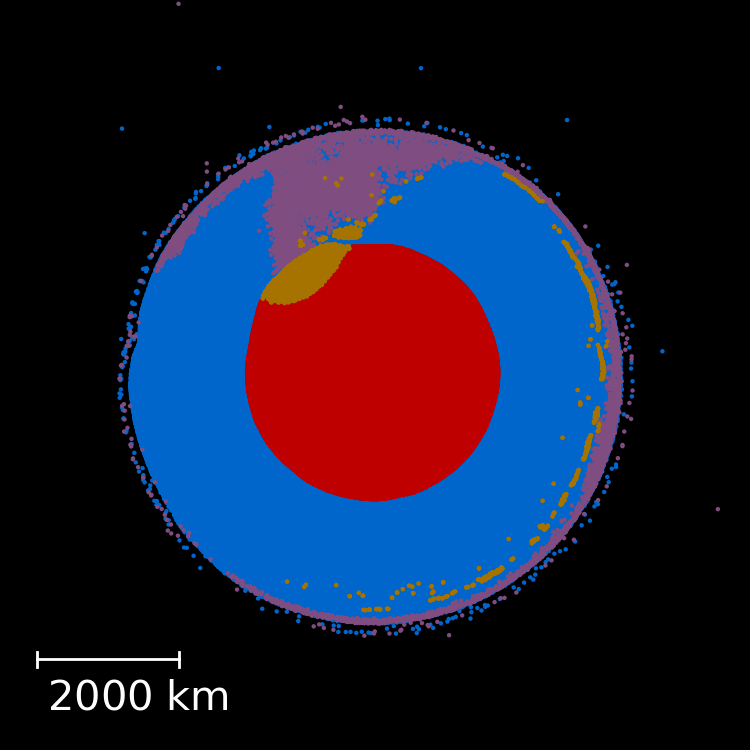}} & \multicolumn{1}{@{}m{1.5in}@{}}{\includegraphics{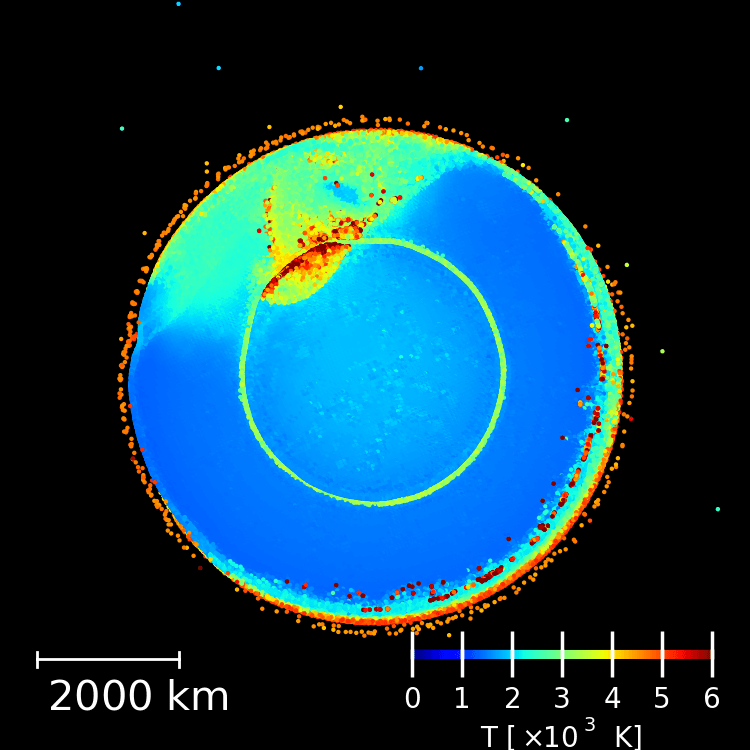}} & \multicolumn{1}{@{}m{1.5in}@{}}{\includegraphics{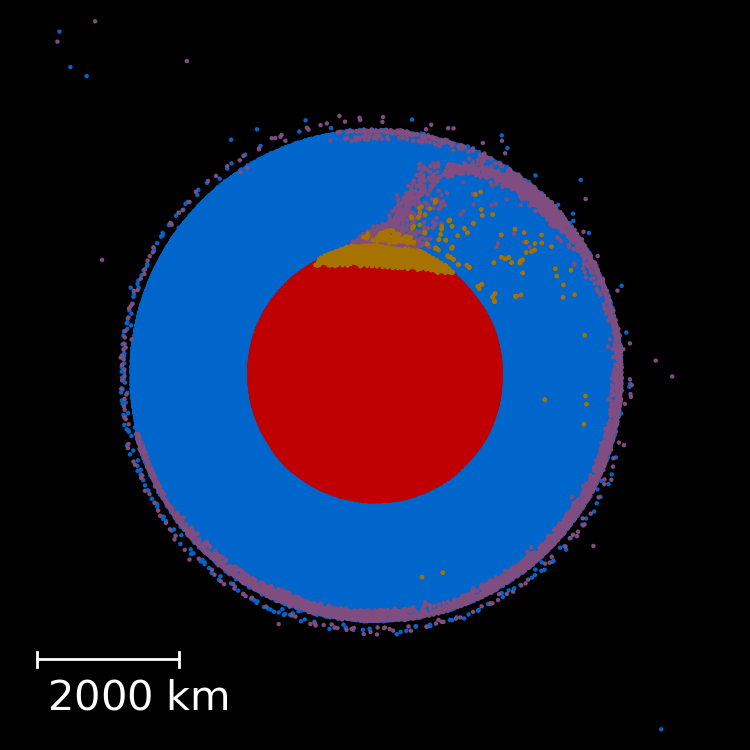}} & \multicolumn{1}{@{}m{1.5in}@{}}{\includegraphics{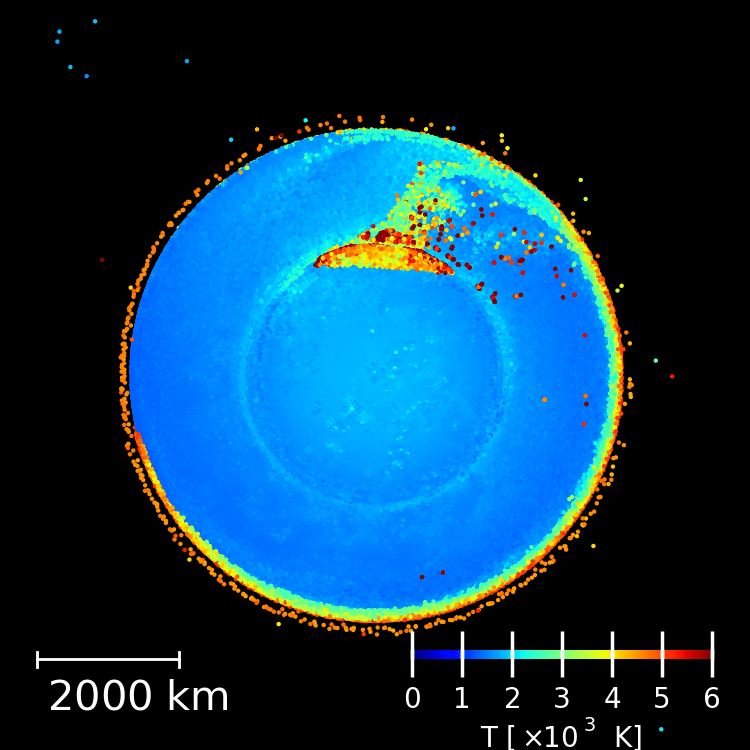}} \\
\rotatebox[origin=c]{90}{\parbox{1.5in}{Tillotson\\density integration}} & \multicolumn{1}{@{}m{1.5in}@{}}{\includegraphics{ao_64800_material}} & \multicolumn{1}{@{}m{1.5in}@{}}{\includegraphics{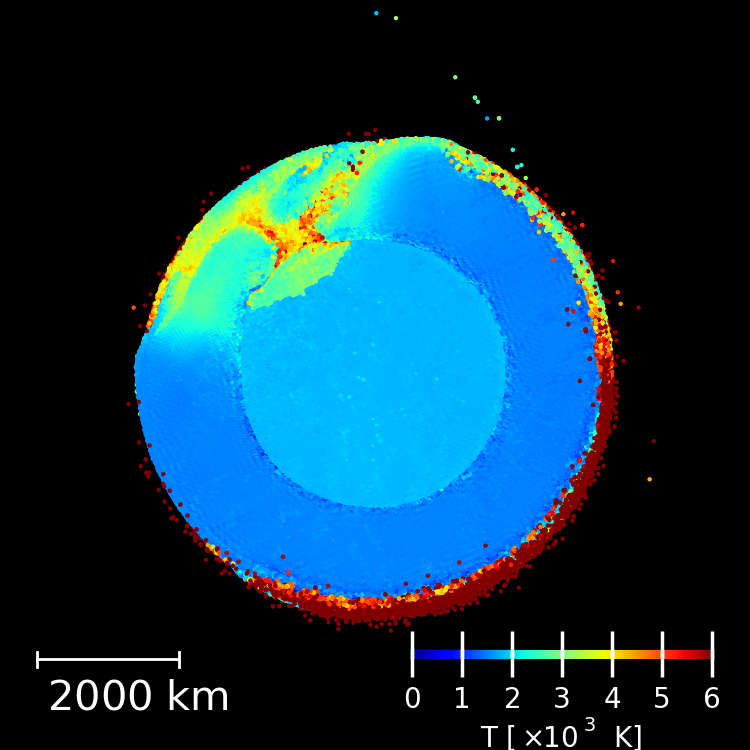}} & \multicolumn{1}{@{}m{1.5in}@{}}{\includegraphics{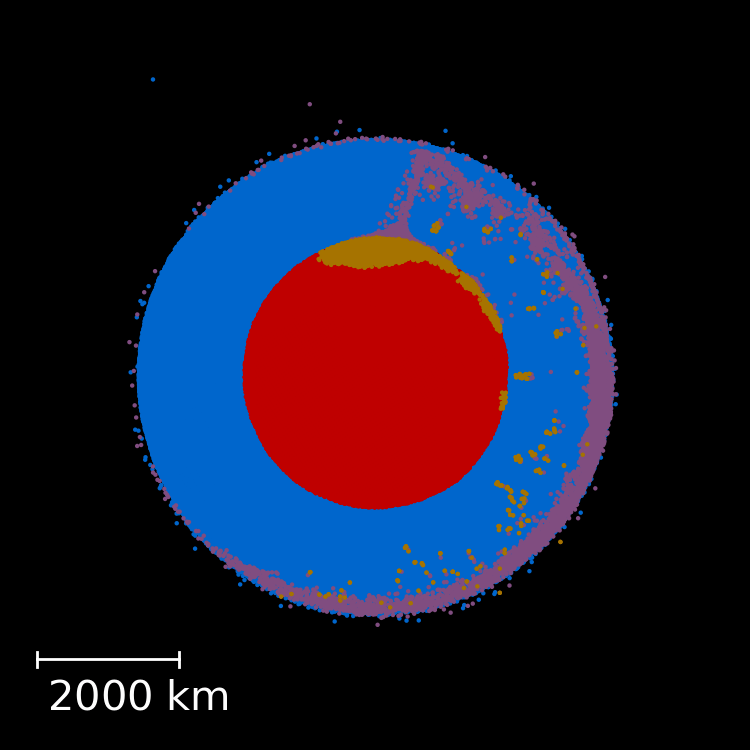}} & \multicolumn{1}{@{}m{1.5in}@{}}{\includegraphics{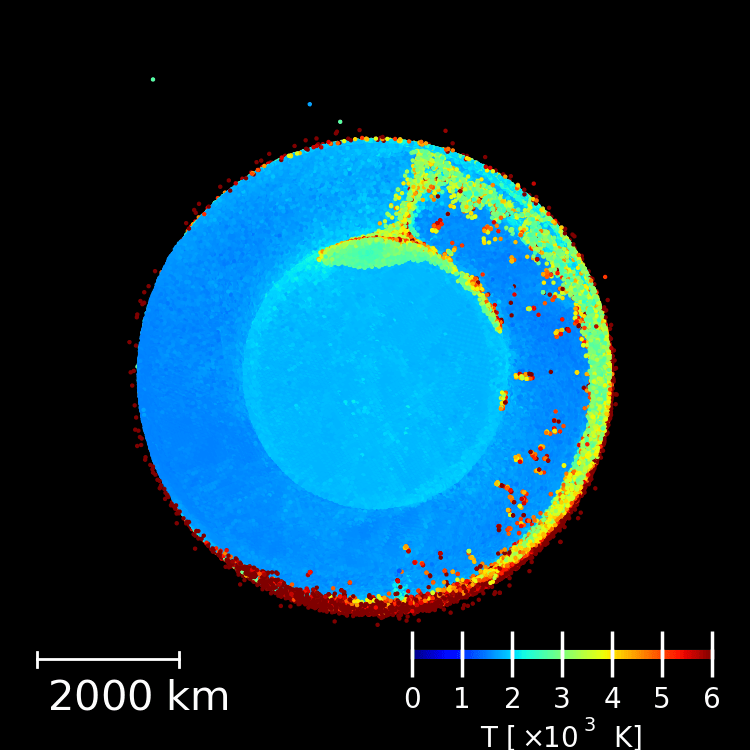}} \\
\rotatebox[origin=c]{90}{\parbox{1.5in}{Tillotson\\density summation}} & \multicolumn{1}{@{}m{1.5in}@{}}{\includegraphics{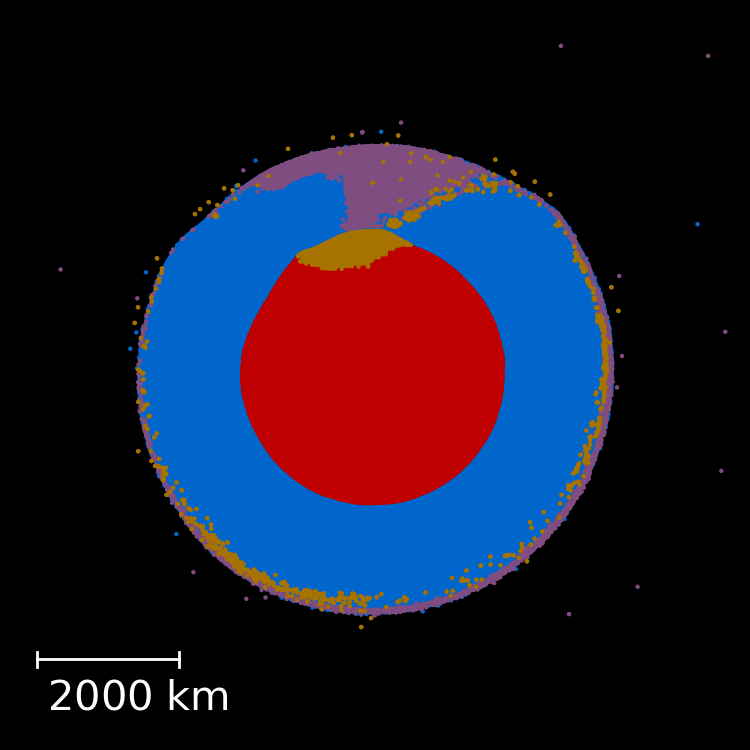}} & \multicolumn{1}{@{}m{1.5in}@{}}{\includegraphics{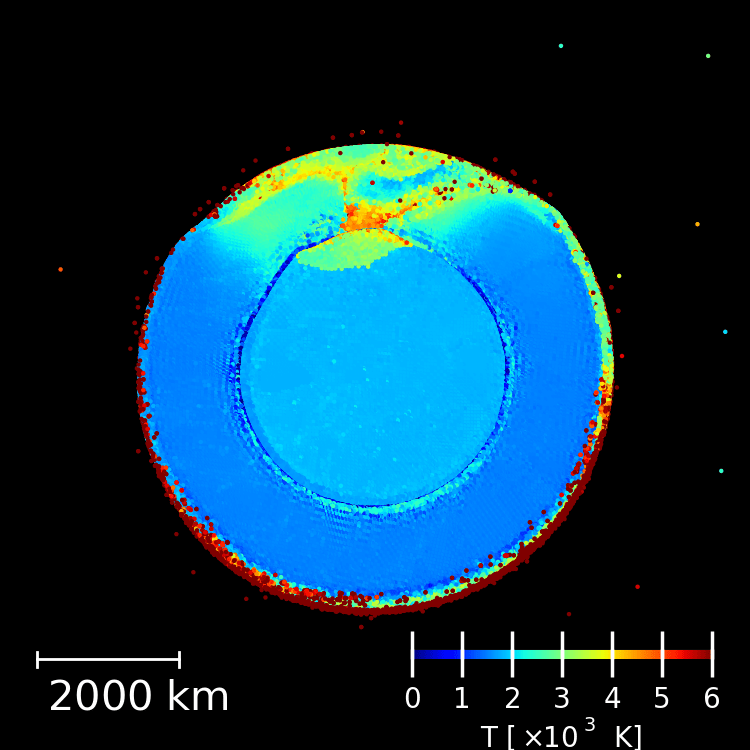}} & \multicolumn{1}{@{}m{1.5in}@{}}{\includegraphics{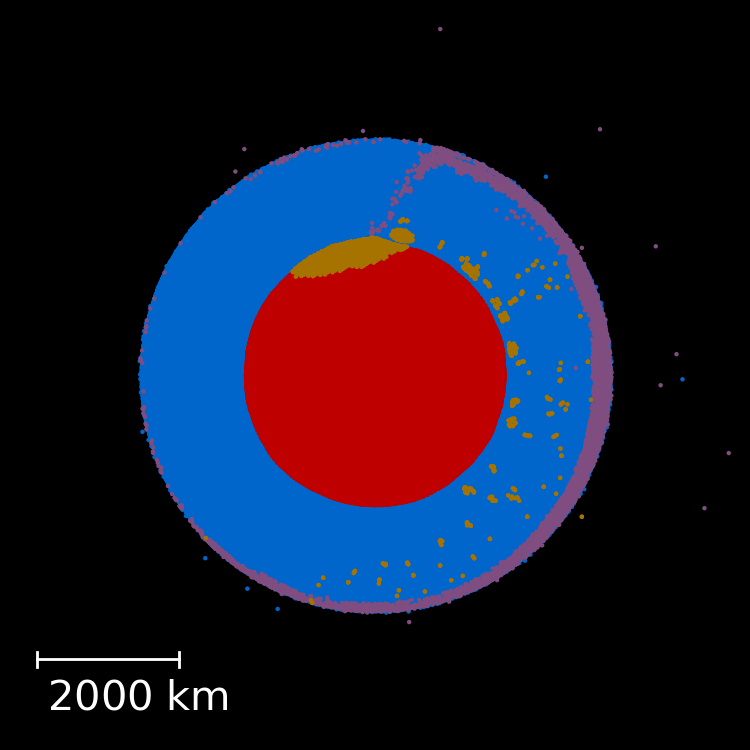}} & \multicolumn{1}{@{}m{1.5in}@{}}{\includegraphics{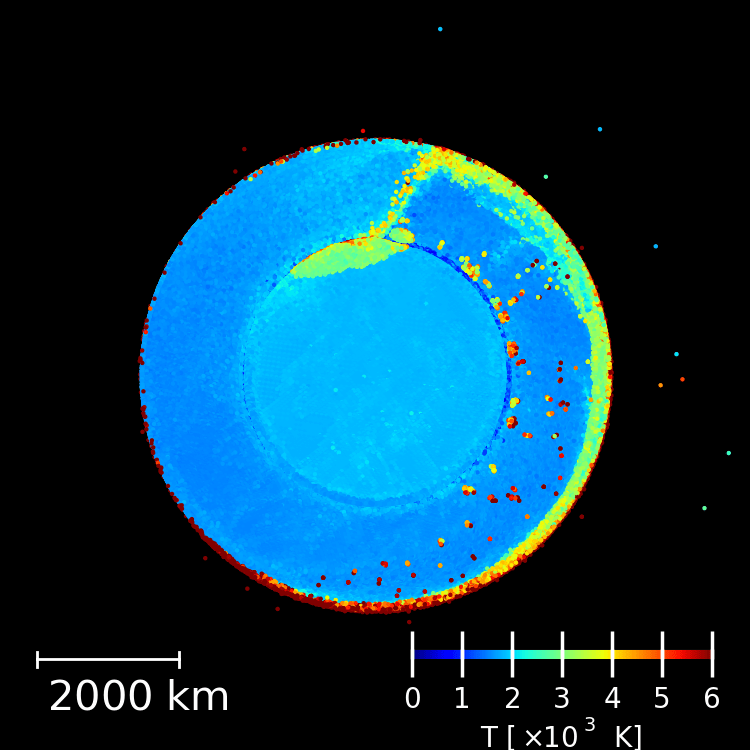}} \\
\end{tabularx}
}
\begin{center}
\caption{Material and temperature plots for the oblique impact cases at the end of the simulations. The top line is the same as the final one of figure~\ref{fig:nomcase:timeseries}.}
\label{fig:comparison:oblique}
\end{center}
\end{figure}

\begin{figure}
\def\arraystretch{0.}
\makebox[\textwidth][c]{
\begin{tabularx}{6.5in}{@{}cc@{}c@{}c@{}c@{}}
 & Solid, material & Solid, temperature & Fluid, material & Fluid, temperature \\
\rotatebox[origin=c]{90}{\parbox{1.5in}{ANEOS\\density integration}} & \multicolumn{1}{@{}m{1.5in}@{}}{\includegraphics{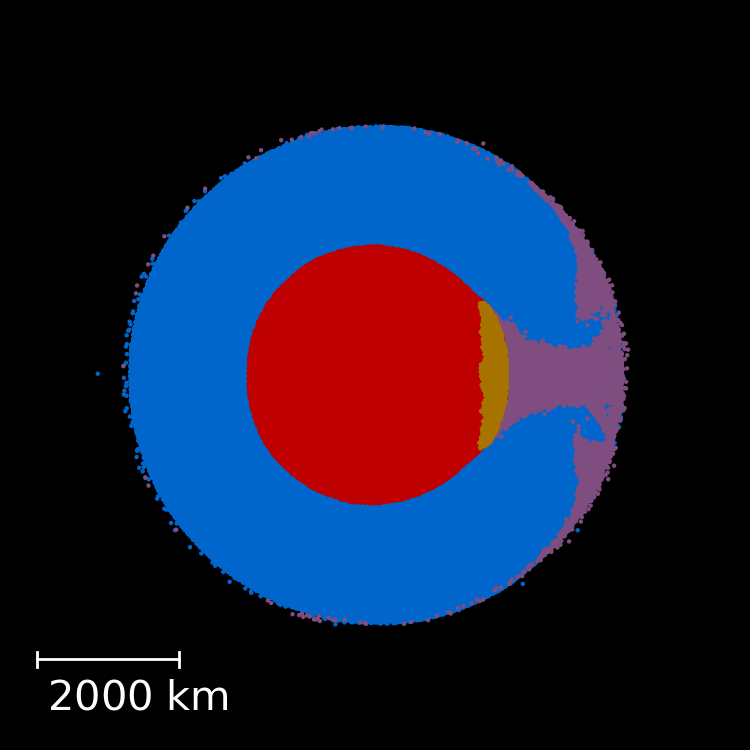}} & \multicolumn{1}{@{}m{1.5in}@{}}{\includegraphics{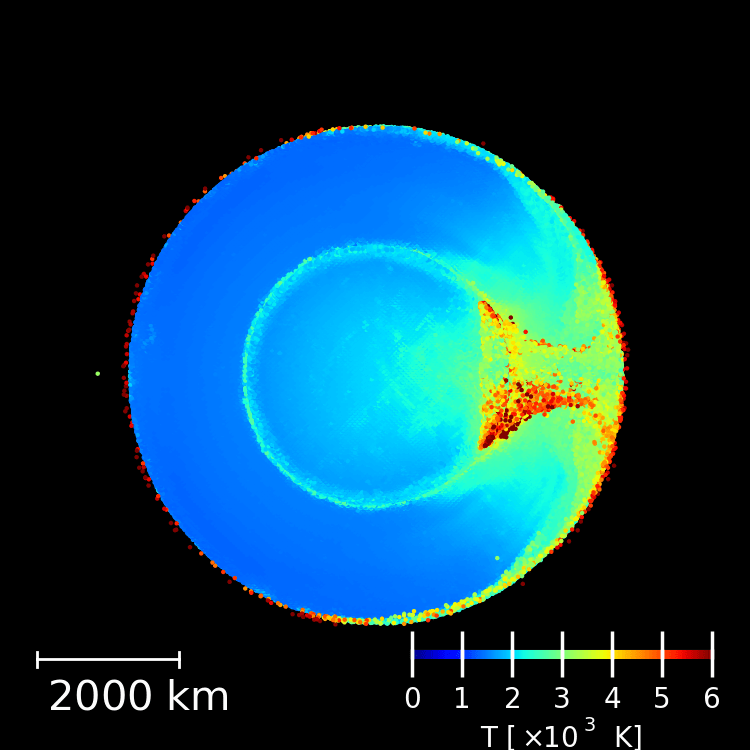}} & \multicolumn{1}{@{}m{1.5in}@{}}{\includegraphics{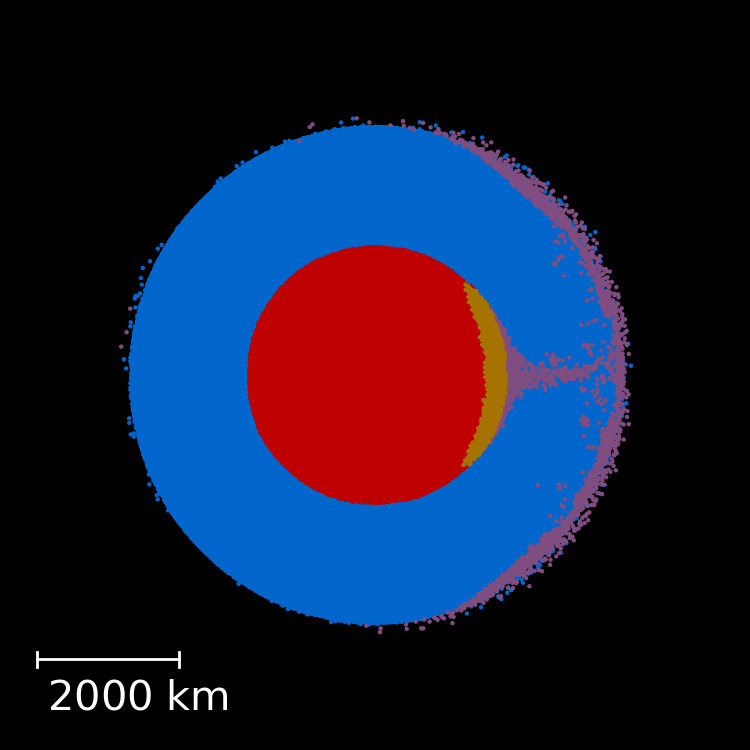}} & \multicolumn{1}{@{}m{1.5in}@{}}{\includegraphics{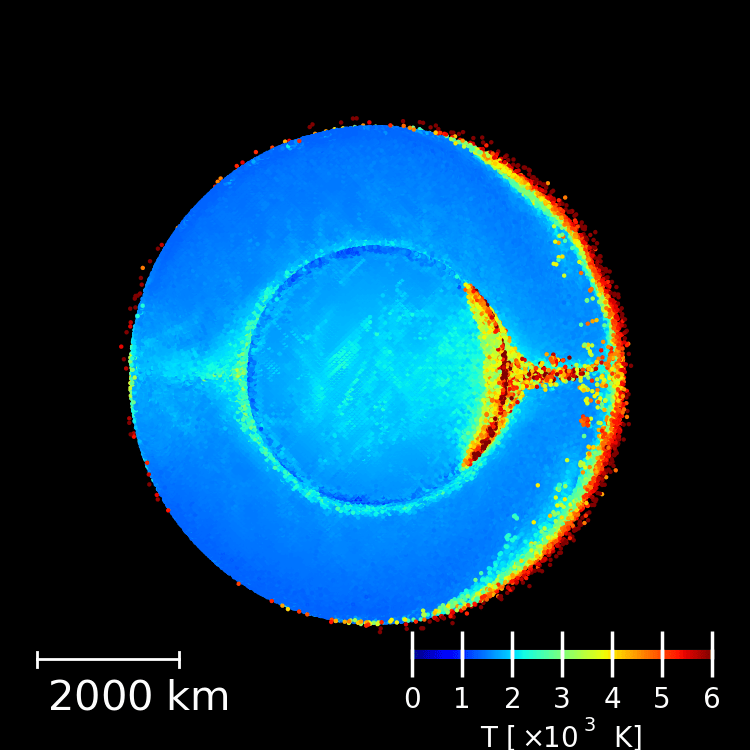}} \\
\rotatebox[origin=c]{90}{\parbox{1.5in}{ANEOS\\density summation}} & \multicolumn{1}{@{}m{1.5in}@{}}{\includegraphics{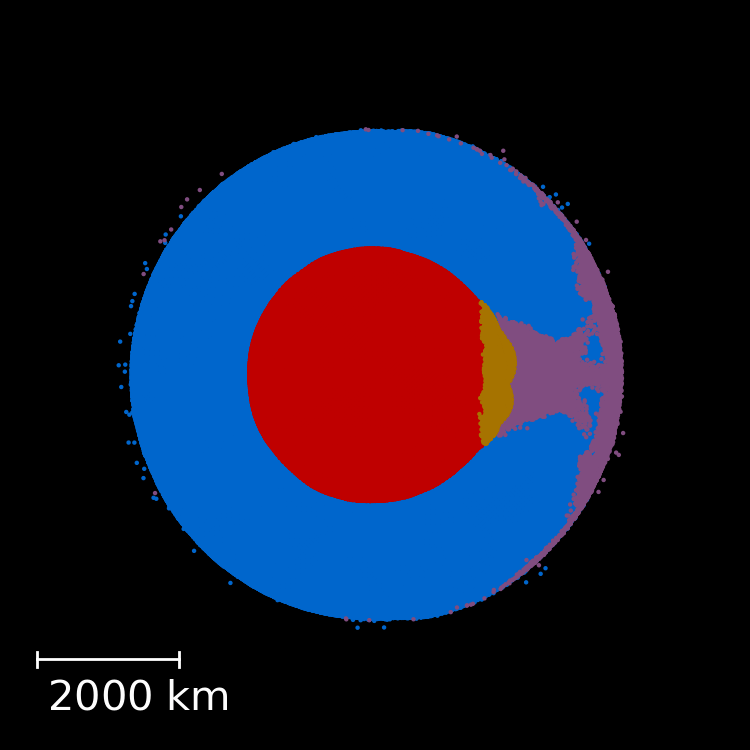}} & \multicolumn{1}{@{}m{1.5in}@{}}{\includegraphics{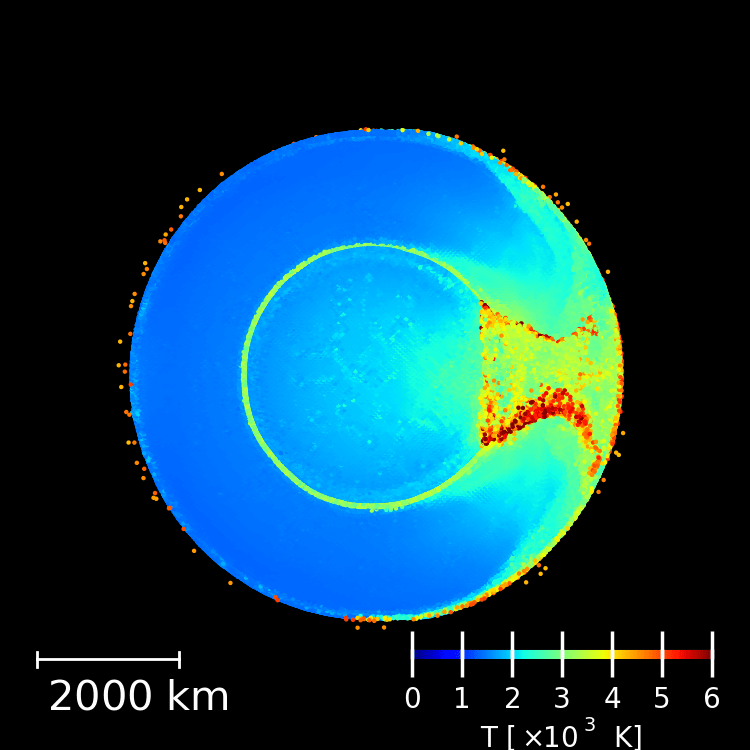}} & \multicolumn{1}{@{}m{1.5in}@{}}{\includegraphics{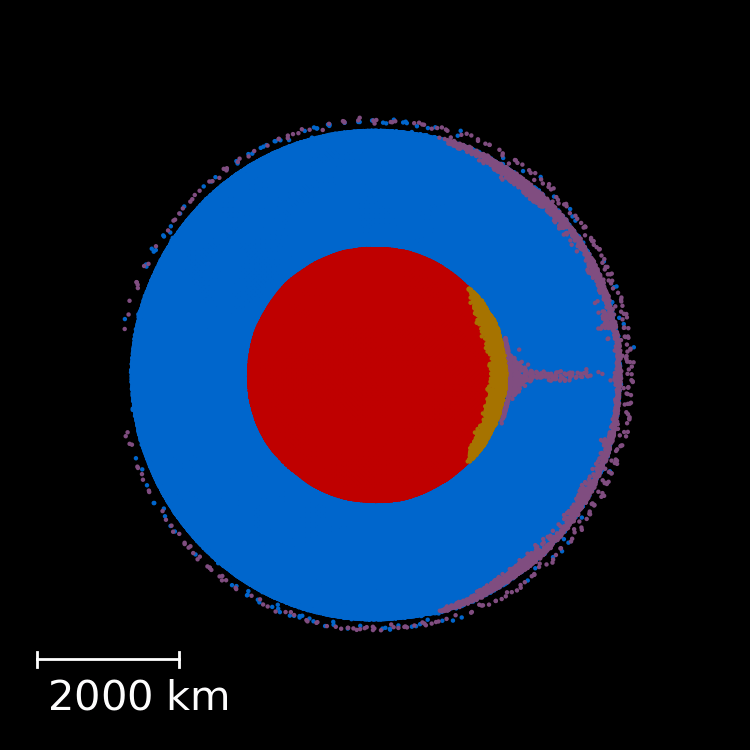}} & \multicolumn{1}{@{}m{1.5in}@{}}{\includegraphics{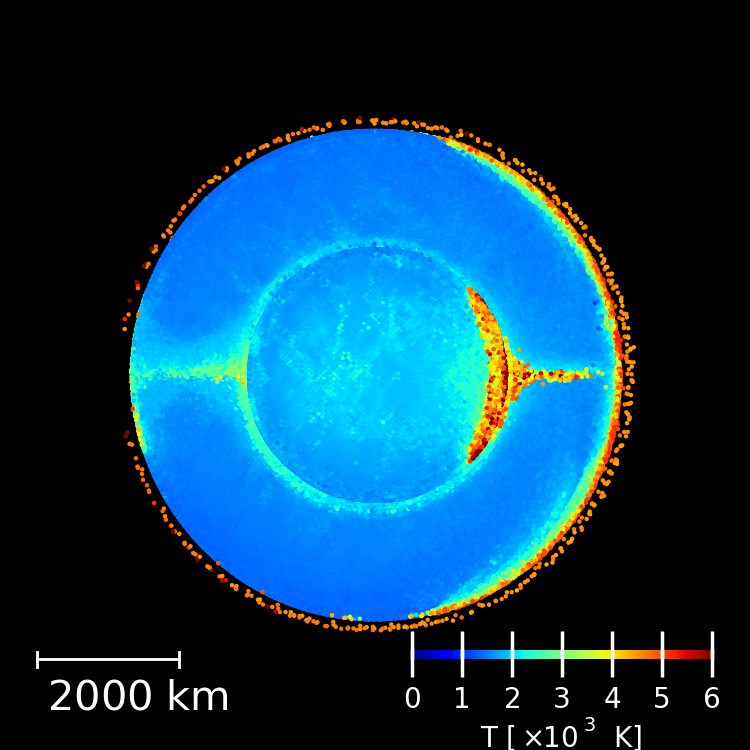}} \\
\rotatebox[origin=c]{90}{\parbox{1.5in}{Tillotson\\density integration}} & \multicolumn{1}{@{}m{1.5in}@{}}{\includegraphics{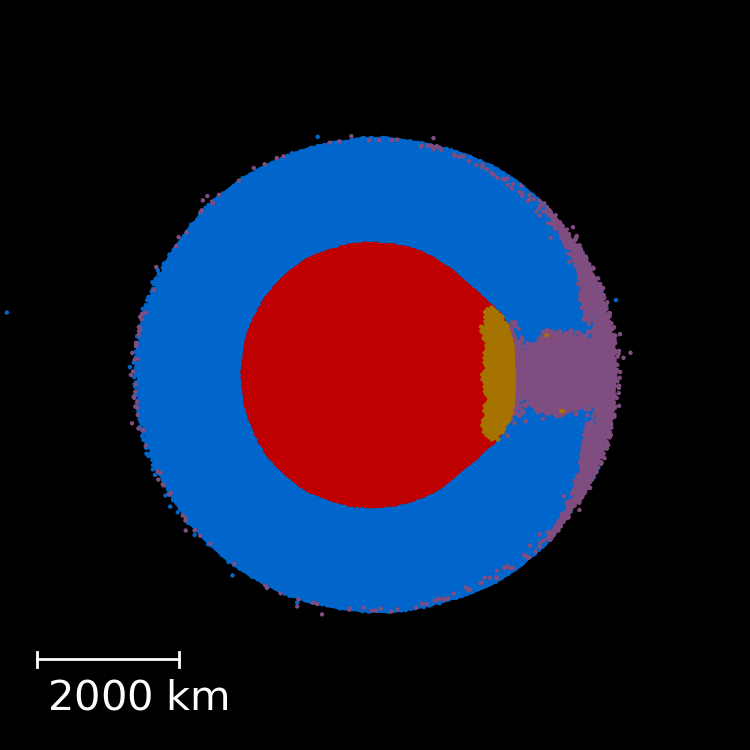}} & \multicolumn{1}{@{}m{1.5in}@{}}{\includegraphics{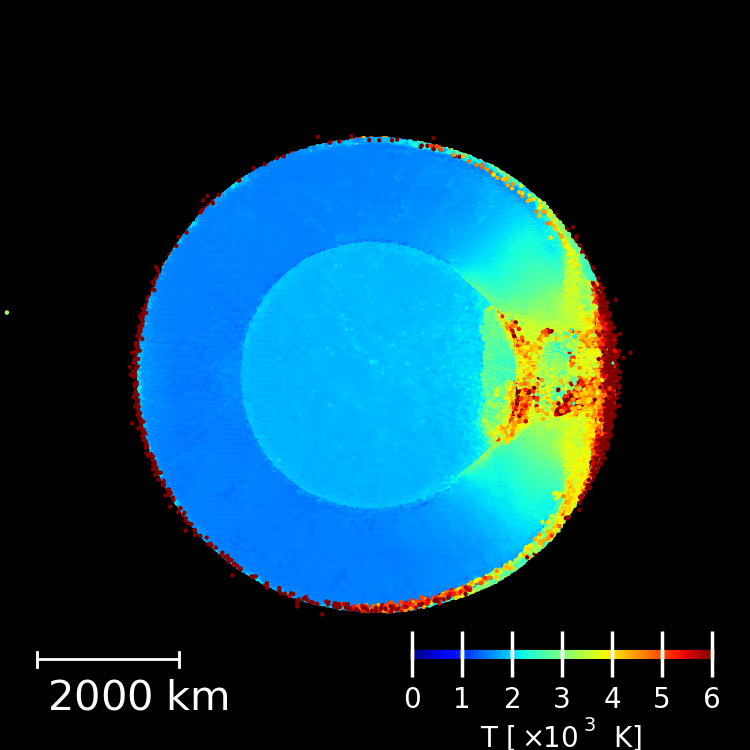}} & \multicolumn{1}{@{}m{1.5in}@{}}{\includegraphics{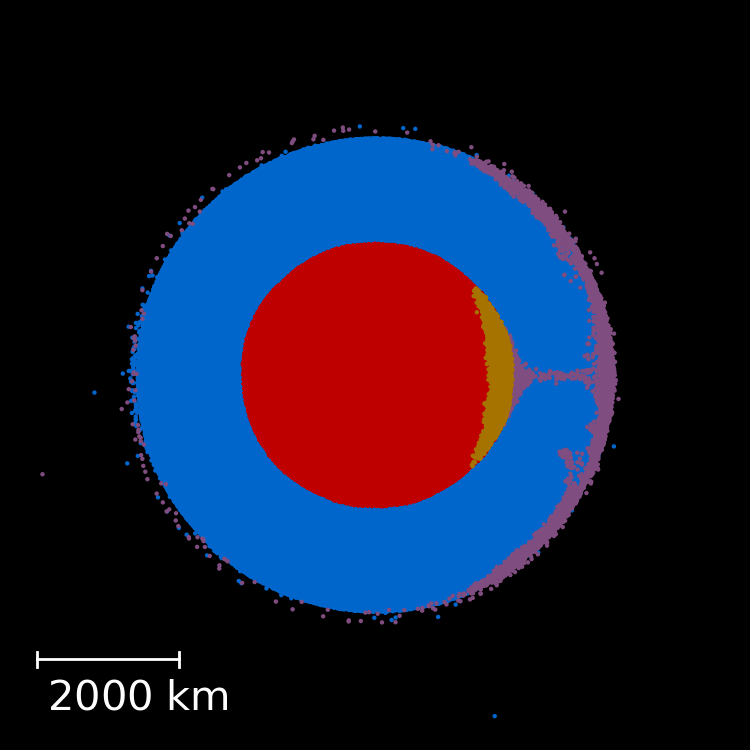}} & \multicolumn{1}{@{}m{1.5in}@{}}{\includegraphics{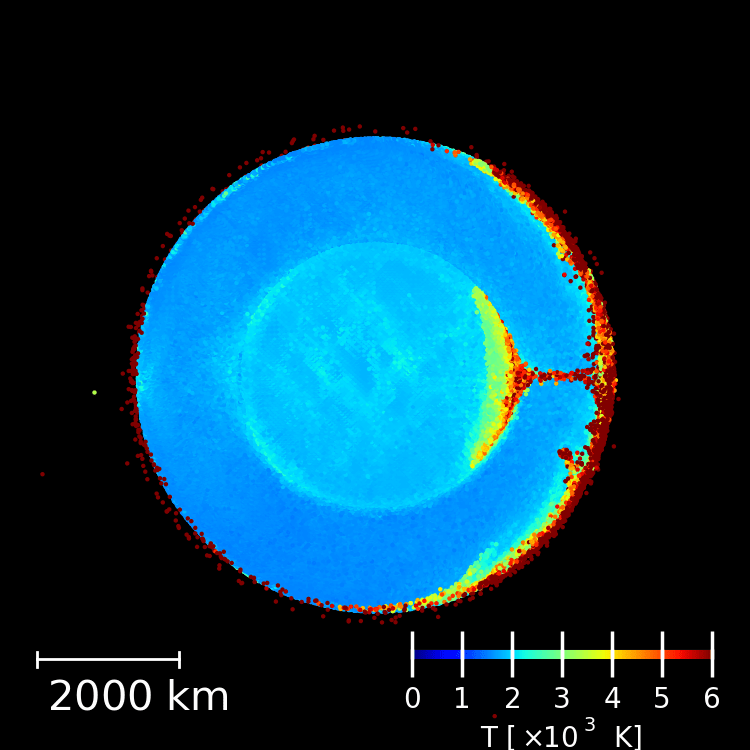}} \\
\rotatebox[origin=c]{90}{\parbox{1.5in}{Tillotson\\density summation}} & \multicolumn{1}{@{}m{1.5in}@{}}{\includegraphics{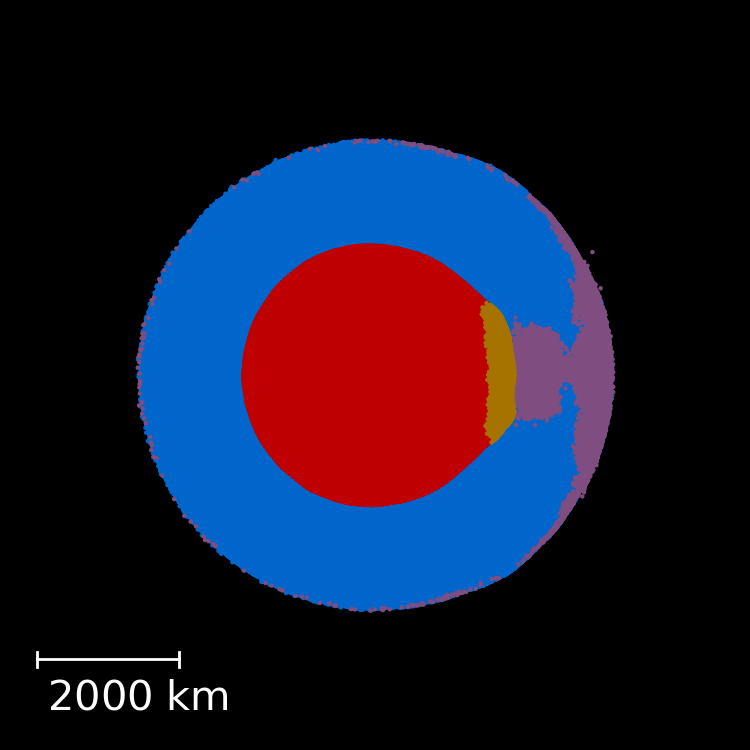}} & \multicolumn{1}{@{}m{1.5in}@{}}{\includegraphics{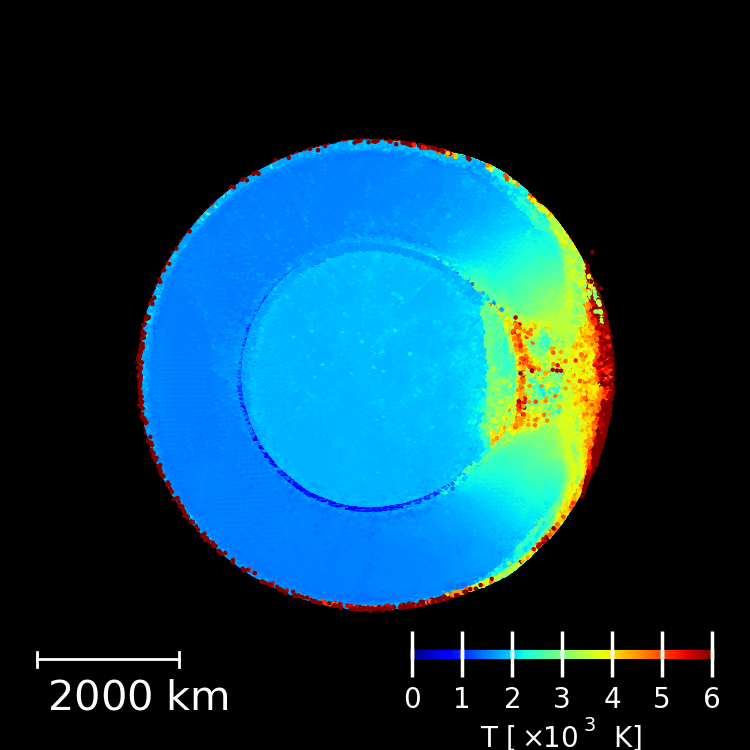}} & \multicolumn{1}{@{}m{1.5in}@{}}{\includegraphics{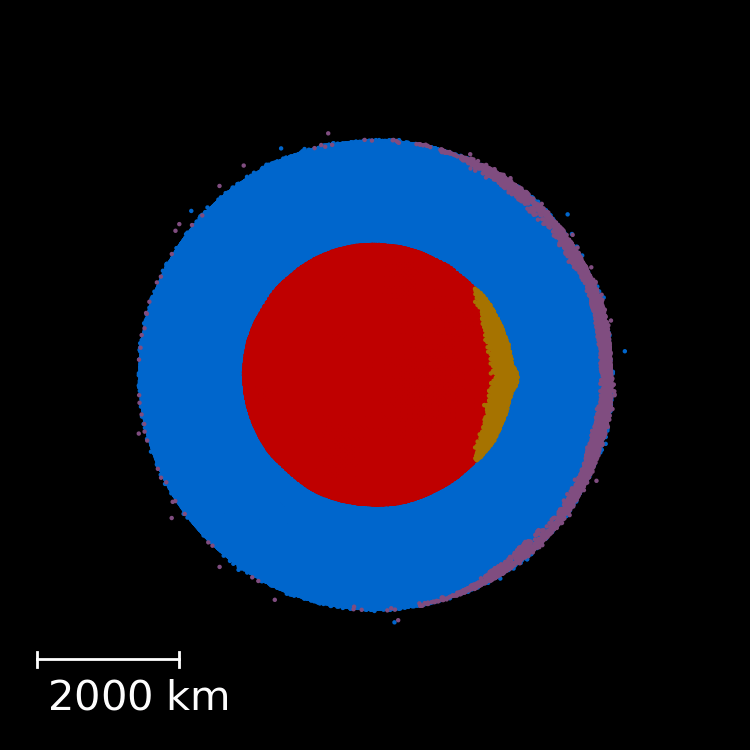}} & \multicolumn{1}{@{}m{1.5in}@{}}{\includegraphics{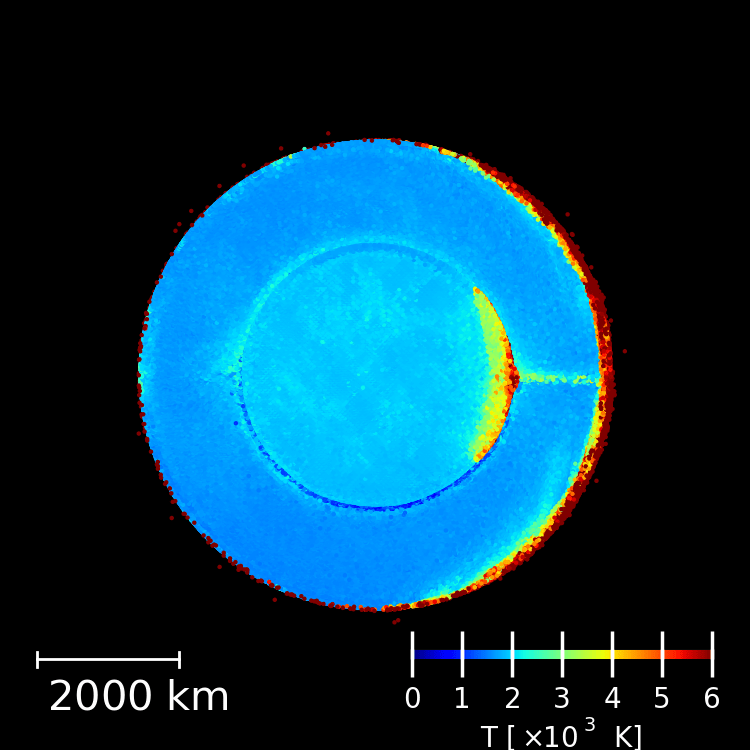}} \\
\end{tabularx}
}
\begin{center}
\caption{Same as figure~\ref{fig:comparison:oblique}, but for the corresponding head-on cases. Initial contact happens on the right side.}
\label{fig:comparison:head-on}
\end{center}
\end{figure}

\begin{figure}
	\begin{center}
		\includegraphics{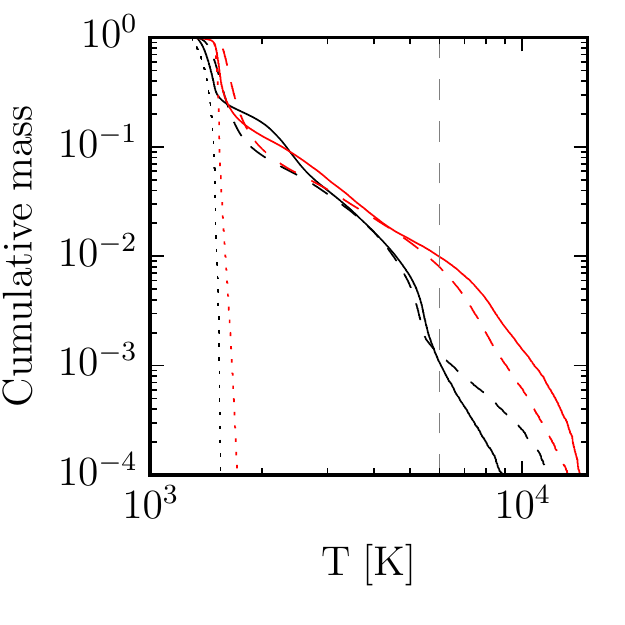}
		\includegraphics{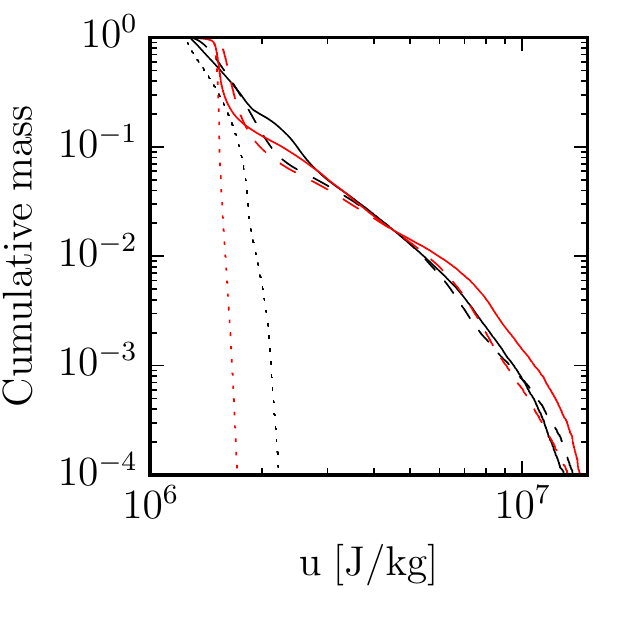}
		\caption{Cumulative mass fraction of the mantle with temperature (left panel) or specific internal energy (right panel) greater than a given value. Black lines denote ANEOS whereas red is for Tillotson. Solid lines are for solid rheology, dashed lines for fluid rheology and the dotted lines show the initial target profile. Only density integration with oblique geometry is considered. The dashed grey line on the temperature plot shows the upper bound of the colour scale on figures~\ref{fig:nomcase:timeseries}, \ref{fig:comparison:oblique} and \ref{fig:comparison:head-on}.}
		\label{fig:temperature:rheo-eos}
	\end{center}
\end{figure}

The results from our series of 16 simulations, using different material rheologies, EoS and numerical schemes are shown in figure~\ref{fig:comparison:oblique} for oblique cases and in figure~\ref{fig:comparison:head-on} for head-on ones. The differences between density integration and summation will be discussed in section \ref{sec:results-density}.
Temperature and energy distributions for four of these cases (oblique impacts; density integration) are provided in figure~\ref{fig:temperature:rheo-eos}.
Note that our nominal case and its fluid counterpart that were discussed in the previous section are shown again in the first row of figure~\ref{fig:comparison:oblique} (ANEOS; integrated density) and with the black lines on figure~\ref{fig:temperature:rheo-eos}.

At first look, we note that the differences (concerning material and temperature distribution) discussed in the previous section between the runs with rheologies  are also present in the simulations using different EoS and numerical techniques (density computation). The more pronounced temperature increase around the impact zone with solid rheology is noticeable on figure~\ref{fig:temperature:rheo-eos} between 2000 and $\unit{3000}{\kelvin}$; it is about $\unit{10-20}{wt\%}$ for ANEOS and a bit less for Tillotson.

The differences between the runs with different EoS are more subtle. For instance, using the Tillotson EoS, the impactor's particles close to the surface have higher temperatures. This effect appears both in figures~\ref{fig:comparison:oblique} and \ref{fig:comparison:head-on} as well as in \ref{fig:temperature:rheo-eos}. In figure \ref{fig:temperature:rheo-eos}, the curves for the two EoS diverge above about $\unit{4000}{\kelvin}$, whereas the energy distribution remains similar.  ANEOS includes the latent heat of vaporisation which our simple conversion formula (eq.~\ref{eq:model:tillotsontemp}) for Tillotson neglects. Hence the high temperatures are overestimated with Tillotson. It should be noted that for silicate, neither EoS take\textit{s} the latent heat of melting into account. One can also see in figure  \ref{fig:temperature:rheo-eos} that the behaviour at low temperature depends on the initial profile, and that almost all mantle gets a small temperature increase (of the order of hundred kelvins).

When looking at head-on cases, again the same general features appear. Hot surface material is now located directly above the impact location. The fluid cases however show some specific results. The impact itself produces surface waves that propagate and are focused at the antipode. At this location, these waves lead to some spallation of mantle material, which gets heated when falling back. We thus see an increase in temperature at the antipode; here there are some notable differences between the Tillotson and the ANEOS cases.

Comparing the influence of both effect, we find that rheology plays an important role for both material and heat distribution, even at this scale. The EoS, despite changing the radius of the bodies by $\unit{\sim4}{\%}$, has little influence on these effects. Compared to Tillotson, the more sophisticated temperature computation by ANEOS, which includes the latent heat of vaporisation, leads to differences only at relatively high temperatures (above $\unit{4000-5000}{\kelvin}$).

\subsection{Numerical effects}

\subsubsection{Rotational instability}

\begin{figure}
	\def\arraystretch{0.}
	\makebox[\textwidth][c]{
		\begin{tabularx}{6in}{@{}c@{}c@{}c@{}c@{}}
			\multicolumn{2}{c}{Corrected} &
			\multicolumn{2}{c}{Non-corrected} \\
			material & temperature & material & temperature \\
			\multicolumn{1}{@{}m{1.5in}@{}}{\includegraphics{ao_64800_material}} &
			\multicolumn{1}{@{}m{1.5in}@{}}{\includegraphics{ao_64800_temperature}} &
			\multicolumn{1}{@{}m{1.5in}@{}}{\includegraphics{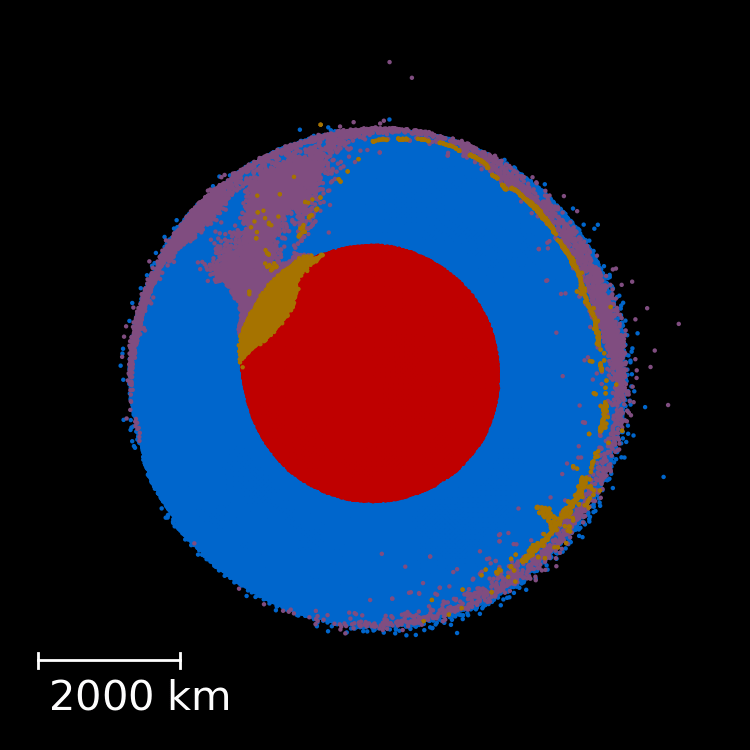}} &
			\multicolumn{1}{@{}m{1.5in}@{}}{\includegraphics{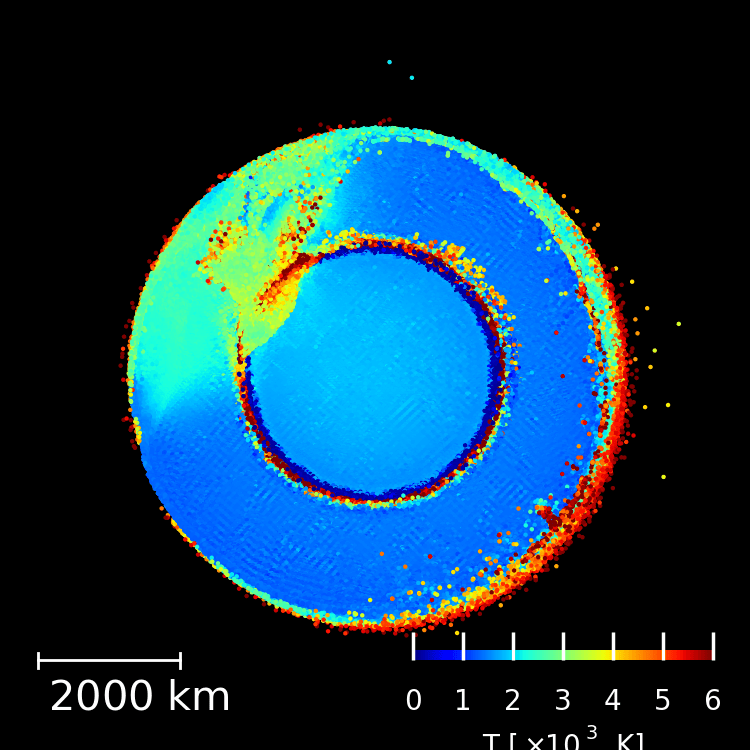}} \\
		\end{tabularx}
	}
	\begin{center}
		\caption{End result comparison for case with angular conservation (two left panels) and without (two right panels). Colours for material plots are the same as in figure~\ref{fig:nomcase:timeseries}. }
		\label{fig:rot:mat}
	\end{center}
\end{figure}

\begin{figure}
\begin{center}
\includegraphics{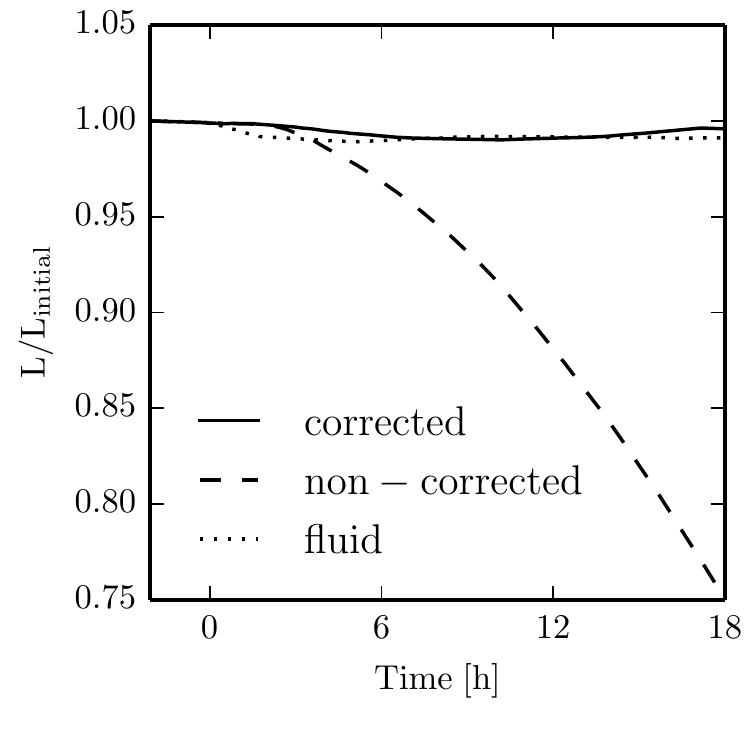}
\caption{Angular momentum as function of time for the two cases shown above ($\unit{45}{\degree}$, ANEOS, solid, integrated density), with/without correction tensor and the corresponding fluid case for comparison.}
\label{fig:rot:angmom-time}
\end{center}
\end{figure}

In figure~\ref{fig:rot:mat} we show the effect of applying the correction tensor in the computation of the stress tensor to avoid rotational instability effects in the case of a solid rheology \citep[e.g.][]{Habil2006Speith}. As it can be seen, the post-impact material and temperature distributions are very similar in the cases with and without correction tensor, except for the temperature artefact due to the set up routine at the core-mantle boundary. We note that the amplitude of the artificial temperature increase at the boundary is much smaller in the simulation with the correction tensor (about $\sim\unit{1000}{\kelvin}$ increase) than in the one without (about $\sim\unit{5000}{\kelvin}$ increase).  Furthermore, the longer term target rotation is slower in the case without correction tensor, because the angular momentum is not conserved.

The angular momentum is plotted as a function of time in figure~\ref{fig:rot:angmom-time} for the two cases. We also include the corresponding fluid case for comparison. For the solid rheology with correction and the fluid rheology, the variation of angular momentum over the whole simulations (about $\unit{20}{\hour}$) is lower than $\unit{1}{\%}$ whereas for the solid rheology without correction, angular momentum is reduced by almost $\unit{25}{\%}$. The decrease starting a few hours after time impact reflects the slow down of the rigid body rotation which is not treated correctly without the correction tensor. Ejected material is not subject to this problem as it does not encounter solid forces. The rotation rate decrease implies that target's position is not the same during late stages of simulations. At 18 hours after impact, the target rotated around 20 degrees less than in the case where angular momentum is conserved (figure \ref{fig:rot:mat}). Here, the target rotates about $\unit{100}{\degree}$ between the merging of both cores at two hours after the impact and the end at 18 hours.

These results confirm that standard SPH does not handle well problems which involve rigid body rotation \citep{Habil2006Speith} and point out the importance of applying the correction tensor to deal with such cases, in particular for situations where the rotation timescale is comparable to the simulated time. 

\subsubsection{Density computation (summation vs. integration)}
\label{sec:results-density}

\begin{figure}
	\begin{center}
		\includegraphics{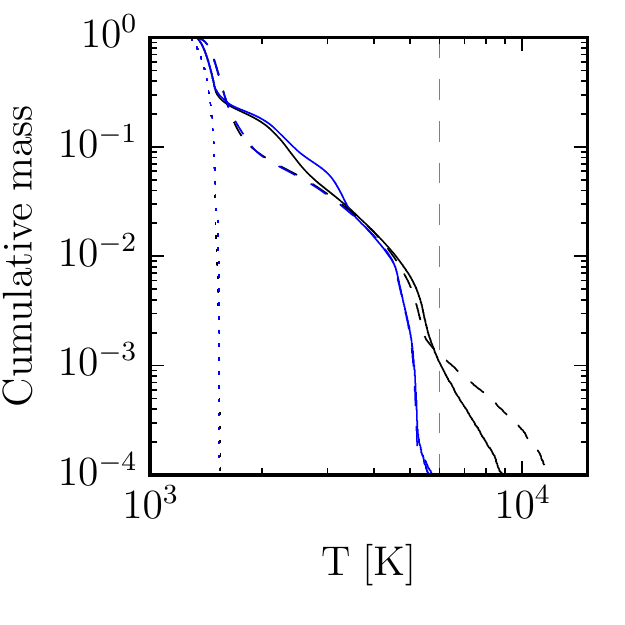}
		\includegraphics{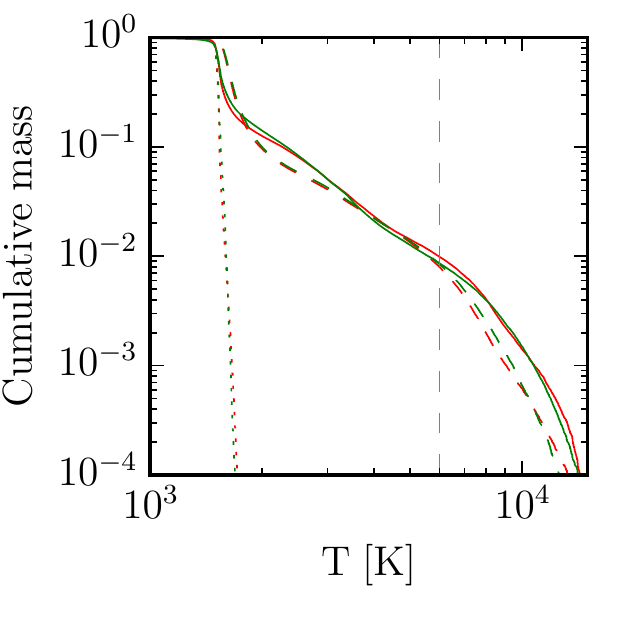}
		\caption{Cumulative mass fraction of mantle with temperature greater than a given value. Same as left panel of figure~\ref{fig:temperature:rheo-eos}, but comparing density computation instead. Left panel is for ANEOS, where blue lines are now density summation. Right panel is Tillotson with green lines for density summation.}
		\label{fig:temperature:dens}
	\end{center}
\end{figure}

As discussed in section \ref{sec:modeling}, we test two different methods to compute the densities of the SPH particles. We use \textit{density integration} for our nominal cases.

Major differences arise at the boundaries between materials or at the surface where steep density gradients are located. For density summation, a group of particles separates in all quantities shown in radial profiles (figures~\ref{fig:profile:impactor} and~\ref{fig:profile:target}). These groups contain particles closer than $2h$ from the boundary and where the adjacent material enters in the SPH sum (or where there is a lack of neighbouring particles when close to the surface). Their density is shifted towards the one beyond the boundary. For integrated density, as no such sum is involved, particles close to the boundary do not separate from the remaining. However, there are still some oscillations close to the discontinuity.

Despite the issues mentioned above, which result in different initial density profiles (at the boundaries), the results at end of the simulations do not show any major differences between the two schemes (figures \ref{fig:comparison:oblique} and \ref{fig:comparison:head-on}). However, there are some subtle effects. For instance, we observe a different behaviour of hot surface particles which were ejected during the early stage of impact and reaccreted later on. With ANEOS and density summation, they form a layer separated from the planet's surface. In the other cases, this feature is less pronounced and not present at all in the cases with density integration. In the ANEOS case, the different densities of the surface particles also affect the temperature (figure~\ref{fig:temperature:dens}, left panel) although the energy distribution remains essentially identical. As surface particle density is lower in the density summation case than for density integration, ANEOS does not find them in the same phase: mixed liquid-vapour for the former and liquid in the latter. Density summation simulations exhibit lower temperature since part of the internal energy is used for the phase transition rather than temperature increase. This explains the absence of material with temperature above $\unit{6000}{\kelvin}$ with ANEOS density summation on the figure. The second small bump located between $\numprint{2000}$ and $\unit{3000}{\kelvin}$ in the case with solid rheology and density summation (solid blue line on figure~\ref{fig:temperature:dens}) is due to the core-mantle boundary. As phase transition does not influence the temperature in the case with  Tillotson EoS, we don't see this feature on the right panel of figure~\ref{fig:temperature:dens}.

We note that the differences between the numerical schemes discussed above result from differences at small scales, representing only a small fraction of the total mass involved in the simulation. Even at the relatively high resolution (one million SPH particles) used in the simulations performed here, some of these small-scale features are under-resolved. This is certainly the case for the layer of reaccreted material discussed above. Also the peak pressures and temperatures produced in the initial stages of the collision may not be fully resolved. 

With the resolution used in the simulations presented here, the numerical scheme with density integration appears to be the better choice as it leads to a better defined surface. It is clear, however, that a more consistent treatment of the boundaries, as well as very high resolution simulations, are required to obtain more accurate physical properties at such small scales.

\subsubsection{Effect of the initial energy profile (constant energy vs. integrated energy)}
\label{sec:tillotson-profile}

\begin{figure}
	\def\arraystretch{0.}
	\makebox[\textwidth][c]{
		\begin{tabularx}{6in}{@{}c@{}c@{}c@{}c@{}}
			\multicolumn{2}{c}{Integrated energy} &
			\multicolumn{2}{c}{Constant energy} \\
			material & temperature & material & temperature \\
			\multicolumn{1}{@{}m{1.5in}@{}}{\includegraphics{uo_64800_material}} &
			\multicolumn{1}{@{}m{1.5in}@{}}{\includegraphics{uo_64800_temperature}} &
			\multicolumn{1}{@{}m{1.5in}@{}}{\includegraphics{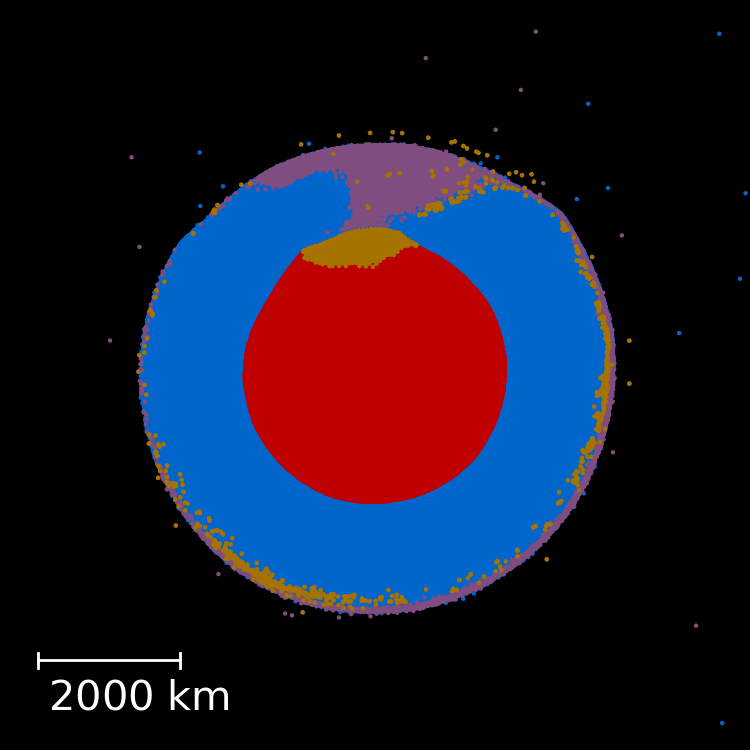}} &
			\multicolumn{1}{@{}m{1.5in}@{}}{\includegraphics{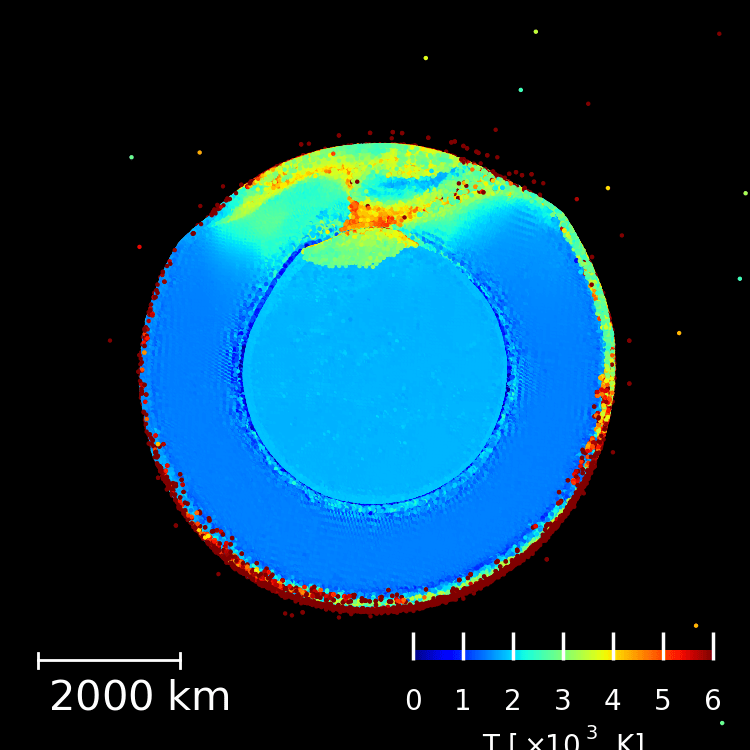}} \\
		\end{tabularx}
	}
	\begin{center}
		\caption{End result comparison for case with evolved internal energy during setup phase (two left panels) and constant (two right panels). Colours for material plots are the same as in figure~\ref{fig:nomcase:timeseries}. }
		\label{fig:tillotson-energy:mat}
	\end{center}
\end{figure}

Bodies computed using the Tillotson EoS were evolved during set up phase with the internal energy allowed to vary. This leads to a spread in temperature for those bodies shown in figures~\ref{fig:profile:impactor} and~\ref{fig:profile:target} as well as the density decrease close to the centre for the impactor computed using density summation.

For test purposes, a second series of bodies was generated with internal energy kept constant. For the latter case we don't observe a density decrease close to body centres.

We don't observe any notable difference between the results from the two different setup schemes. A comparison is shown in figure~\ref{fig:tillotson-energy:mat}. The main differences are located at the core-mantle boundary where the effect of the set up method discussed here and the instability with solid rheology explained in section \ref{sec:results-nomcase} overlap. This effect is small compared with all the previously discussed parameters.

\subsection{Rotation period}

\begin{table}
	\begin{center}
		\begin{tabular}{llll}
			\hline
			EoS & Rheology & Density & Period [Sidereal day]  \\
			\hline
			\multirow{4}{*}{ANEOS} & Solid & Integrated & 1.22 \\
			& Solid & Summed & 1.19 \\
			& Fluid & Integrated & 1.11 \\
			& Fluid & Summed & 1.14 \\
			\hline
			\multirow{4}{*}{Tillotson} & Solid & Integrated & 1.18 \\
			& Solid & Summed & 0.95 \\
			& Fluid & Integrated & 1.13 \\
			& Fluid & Summed & 1.13 \\
			\hline
		\end{tabular}
	
		\caption{Impact-induced rotation periods for oblique geometry runs}
		\label{tab:rotperiod}
	\end{center}
\end{table}

The rotation periods resulting from the oblique impacts considered in this study, calculated by applying a correction tensor to the angular momentum, are presented in table~\ref{tab:rotperiod}. They are comparable to Mars' present rotation period.  All fluid rheology runs have a very similar rotation periods ($\unit{1.11-1.14}{Martian~sideral~days}$). The solid rheology usually leads to a slightly slower rotation, except for the case with Tillotson EoS and density summation which does not follow this trend and is rotating faster.

\subsection{Ejecta, disc mass and escaping material}

\begin{table}
	\begin{center}
		\begin{tabular}{llllll}
			\hline
			& & & Ejecta & Disc & Escaping \\
			EoS & Rheology & Density & [$\mathrm{M_{Mars}}$] & [$\mathrm{M_{Mars}}$] & [$\mathrm{M_{Mars}}$] \\
			\hline
			\multirow{4}{*}{ANEOS} & Solid & Integrated  & $1.4\cdot10^{-3}$ & $2.2\cdot10^{-5}$ & $2.7\cdot10^{-3}$ \\
			& Solid & Summed & $1.3\cdot10^{-3}$ & $2.2\cdot10^{-4}$ & $2.5\cdot10^{-3}$ \\
			& Fluid & Integrated & $1.3\cdot10^{-3}$ & 0 & $2.0\cdot10^{-3}$ \\
			& Fluid & Summed & $1.4\cdot10^{-3}$ & $2.1\cdot10^{-4}$ & $1.9\cdot10^{-3}$ \\
			\hline
			\multirow{4}{*}{Tillotson} &  Solid & Integrated & $1.6\cdot10^{-3}$ & $5.6\cdot10^{-5}$ & $2.8\cdot10^{-3}$ \\
			& Solid & Summed & $1.5\cdot10^{-3}$ & $1.8\cdot10^{-4}$ & $2.6\cdot10^{-3}$ \\
			& Fluid & Integrated & $1.2\cdot10^{-3}$ & 0 & $2.0\cdot10^{-3}$ \\
			& Fluid & Summed & $1.3\cdot10^{-3}$ & $1.0\cdot10^{-4}$ & $1.8\cdot10^{-3}$ \\
			\hline
		\end{tabular}
		
		\caption{Non accreted masses for oblique geometry runs. \textit{Ejecta} column gives the amount of bound ejecta, computed at 6 hours after impact. \textit{Disc} is the amount of material that remains orbiting around the simulations' end (18 hours after impact) and \textit{Escaping} represents the unbound material, also at simulations' end.}
		\label{tab:ejecta}
	\end{center}
\end{table}

We provide in table~\ref{tab:ejecta} components of non-accreted material. The amount of unbound material ("Escaping" column) remains almost constant from several hours after impact onward. At the end of the simulations, it represents the biggest component of the ejecta. We find that runs with fluid rheology lead to roughly 30\% less escaping mass than the runs with solid rheology. This is visible in figure~\ref{fig:cumulmass:canonical} as the impactor's mantle (purple lines) is less present in the planet with solid rheology. The impactor penetrate less deeply inside the target during the initial phase for the latter case. As consequence, there is a slight increase in the amount of material not directly accreted by the planet, which is reflected in the escaping mass.

The amount of material which is still bound but not (yet) reaccreted is lower than the amount of escaping material by 6 hours after impact. It decreases as material fall back onto the planet and eventually reaches a value about three times lower at the end. For a further analysis of that part, the reader is referred to \citet{Companion}.

Also shown in table~\ref{tab:ejecta} are the disc's masses. These values fluctuate and tend to increase with time. Particles composing the disc come from a small subset of the bound ejecta. We find large variations of the disc mass, depending mostly on the numerical scheme. However, the discs masses are very small (a few 10$^{-4}$ $M_{Mars}$ at most) and are under-resolved (they contain between 138  and 332 SPH particles for density summation and up to 74 for density integration). Much higher resolution would be required to obtain meaningful results in terms of the properties of such small discs. We note that recent studies of the impact-formation of the Martian moons \citep{2015IcarusCitron,2016LPICanupSalmon}, which considered comparable disc masses, used similar or lower resolution and did not include material strength. 

\section{Discussion and Conclusions}

In this paper, we model large-scale ($\approx\unit{2000}{\kilo\meter}$) impacts on a Mars-like planet using an updated SPH code \citep{2012IcarusReufer}, which includes self-gravity, a newly implemented strength model \citep{2015PSSJutzi} and various EoS (Tilloston and M-ANEOS). A subset of these simulations is used as initial conditions in a thermochemical code to study the long-term interior evolution of Mars \citep{Companion}. Here, we investigate the effects of material strength and of using different EoS. Some potential numerical effects (in the context of discontinuities) which may result from the different ways to compute the density in SPH  are studied as well. Finally, an advanced SPH scheme to avoid rotational instability in rigid body rotation \citep{Habil2006Speith} is applied and tested in the regime of large-scale collisions.

We find that in the collision regime considered here (with impact velocities of $\unit{4}{\kilo\meter\per\second}$), the Tillotson and ANEOS equations of state lead to post-impact temperature distributions which are quite similar in the temperature range below $\unit{\sim4000}{\kelvin}$ with only subtle differences. This indicates that the Tillotson EoS works reasonably well in this regime (where vaporisation is not significant) and the simple estimation of the temperature from the specific energy by using a constant heat capacity $T=u/c_p$ is justified.

On the other hand, our results strongly suggest that the effect of material strength is substantial even for the large-scale collisions considered here. When strength is taken into account, the post-impact distributions of the impactor material and temperature are very different. For instance, in this case, the heat generated by the impact is spread out over a much broader region of mantle material localised around the impact point.
We expect this finding to be generally important  for collisions in this regime. We note that previous SPH simulations at this scale \citep[e.g.][]{2008NatureMarinova, 2005ScienceCanup,2015IcarusCitron,2016LPICanupSalmon} did not include material strength. Our results also suggested that  Pluto-Charon type collisions \citep[e.g.][]{2005ScienceCanup} might well be in a regime where material strength is not negligible.  

The different density computation schemes used here do not show large differences in the final global outcome of the simulations. However, we observe some differences at small scales, and a more consistent treatment of density discontinuities \citep[e.g.][]{2013PASJHosonoSaitohMakino} is needed. Our study confirms that the known issue of standard SPH in the case of rigid body rotation is solved by using an appropriate correction tensor which increases the SPH consistency \citep{Habil2006Speith}. 

In conclusion, our results show that in the modelling of collisions at scales of $\approx\unit{2000}{\kilo\meter}$, it is essential to use a material strength model.  The details of the EOS are less important in this regime. Finally, an improved treatment of boundaries as well as a very high resolution are required to obtain accurate physical properties at small scales. 

\appendix
\section*{Acknowledgements}

We thank Gregor~J.~Golabek and Taras~V.~Gerya for helpful discussion in the study's design, and anonymous referee for suggestions that improved this paper. A.E. acknowledges the financial support of the Swiss National Science Foundation under grant 200020\_17246. This work has been carried out within the frame of the National Centre for Competence in Research PlanetS supported by the Swiss National Science Foundation. The authors acknowledge the financial support of the SNSF.

\bibliographystyle{hplainnat}
\bibliography{mars_bib}

\begin{thebibliography}{40}
\providecommand{\natexlab}[1]{#1}
\providecommand{\url}[1]{\texttt{#1}}
\expandafter\ifx\csname urlstyle\endcsname\relax
  \providecommand{\doi}[1]{doi: #1}\else
  \providecommand{\doi}{doi: \begingroup \urlstyle{rm}\Url}\fi

\bibitem[{Andrews-Hanna} et~al.(2008){Andrews-Hanna}, {Zuber}, and
  {Banerdt}]{2008NatureAndrewsHanna}
J.~C. {Andrews-Hanna}, M.~T. {Zuber}, and W.~B. {Banerdt}.
\newblock {The Borealis basin and the origin of the martian crustal dichotomy}.
\newblock \emph{Nature}, 453:\penalty0 1212--1215, June 2008.
\newblock \doi{10.1038/nature07011}.

\bibitem[{Asphaug}(2010)]{2010ChEGAsphaug}
E.~{Asphaug}.
\newblock {Similar-sized collisions and the diversity of planets}.
\newblock \emph{Chemie der Erde / Geochemistry}, 70:\penalty0 199--219, 2010.
\newblock \doi{10.1016/j.chemer.2010.01.004}.

\bibitem[{Asphaug} and {Reufer}(2014)]{2014AGUFMAsphaugReufer}
E.~I. {Asphaug} and A.~{Reufer}.
\newblock {Making an Iron Planet: The Case for Repeated Hit and Run
  Collisions}.
\newblock \emph{AGU Fall Meeting Abstracts}, page~A3, December 2014.

\bibitem[{Barnes} and {Hut}(1986)]{1986NatureBarnesHut}
J.~{Barnes} and P.~{Hut}.
\newblock {A hierarchical O(N log N) force-calculation algorithm}.
\newblock \emph{Nature}, 324:\penalty0 446--449, December 1986.
\newblock \doi{10.1038/324446a0}.

\bibitem[Benz(1991)]{1989BookBenz}
W.~Benz.
\newblock An introduction to computation methods in hydrodynamics.
\newblock In C.~B. de~Loore, editor, \emph{Late stages of stellar evolution,
  computational methods in astrophysical hydrodynamics}, volume 373, pages
  258--312. Springer, 1991.
\newblock ISBN 0387536205.

\bibitem[{Benz} and {Asphaug}(1994)]{1994IcarusBenzAsphaug}
W.~{Benz} and E.~{Asphaug}.
\newblock {Impact simulations with fracture. I - Method and tests}.
\newblock \emph{Icarus}, 107:\penalty0 98, January 1994.
\newblock \doi{10.1006/icar.1994.1009}.

\bibitem[{Benz} and {Asphaug}(1995)]{1995IcarusBenzAsphaug}
W.~{Benz} and E.~{Asphaug}.
\newblock {Simulations of brittle solids using smooth particle hydrodynamics}.
\newblock \emph{Comput. Phys. Commun.}, 1-2:\penalty0 253--265, 1995.
\newblock \doi{10.1016/0010-4655(94)00176-3}.

\bibitem[{Benz} et~al.(1988){Benz}, {Slattery}, and
  {Cameron}]{1988IcarusBenzSlatteryCameron}
W.~{Benz}, W.~L. {Slattery}, and A.~G.~W. {Cameron}.
\newblock {Collisional stripping of Mercury's mantle}.
\newblock \emph{Icarus}, 74:\penalty0 516--528, June 1988.
\newblock \doi{10.1016/0019-1035(88)90118-2}.

\bibitem[{Benz} et~al.(2007){Benz}, {Anic}, {Horner}, and
  {Whitby}]{2007SSRvBenzAnicHornerWhitby}
W.~{Benz}, A.~{Anic}, J.~{Horner}, and J.~A. {Whitby}.
\newblock {The Origin of Mercury}.
\newblock \emph{Space Sci. Rev.}, 132:\penalty0 189--202, October 2007.
\newblock \doi{10.1007/s11214-007-9284-1}.

\bibitem[{Cameron} and {Ward}(1976)]{1976LPICameronWard}
A.~G.~W. {Cameron} and W.~R. {Ward}.
\newblock {The Origin of the Moon}.
\newblock In \emph{{Lunar and Planetary Science Conference}}, volume~7 of
  \emph{{Lunar and Planetary Science Conference}}, page 120, March 1976.

\bibitem[{Canup}(2004)]{2004IcarusCanup}
R.~M. {Canup}.
\newblock {Simulations of a late lunar-forming impact}.
\newblock \emph{Icarus}, 168:\penalty0 433--456, April 2004.
\newblock \doi{10.1016/j.icarus.2003.09.028}.

\bibitem[{Canup}(2005)]{2005ScienceCanup}
R.~M. {Canup}.
\newblock {A Giant Impact Origin of Pluto-Charon}.
\newblock \emph{Science}, 307:\penalty0 546--550, January 2005.
\newblock \doi{10.1126/science.1106818}.

\bibitem[{Canup}(2011)]{2011AJCanup}
R.~M. {Canup}.
\newblock {On a Giant Impact Origin of Charon, Nix, and Hydra}.
\newblock \emph{Astron. J.}, 141:\penalty0 35, February 2011.
\newblock \doi{10.1088/0004-6256/141/2/35}.

\bibitem[{Canup}(2012)]{2012ScienceCanup}
R.~M. {Canup}.
\newblock {Forming a Moon with an Earth-like Composition via a Giant Impact}.
\newblock \emph{Science}, 338:\penalty0 1052--, November 2012.
\newblock \doi{10.1126/science.1226073}.

\bibitem[{Canup} and {Asphaug}(2001)]{2001NatureCanupAsphaug}
R.~M. {Canup} and E.~{Asphaug}.
\newblock {Origin of the Moon in a giant impact near the end of the Earth's
  formation}.
\newblock \emph{Nature}, 412:\penalty0 708--712, August 2001.

\bibitem[{Canup} and {Salmon}(2016)]{2016LPICanupSalmon}
R.~M. {Canup} and J.~{Salmon}.
\newblock {On an Origin of Phobos-Deimos by Giant Impact}.
\newblock In \emph{Lunar and Planetary Science Conference}, volume~47 of
  \emph{Lunar and Planetary Science Conference}, page 2598, March 2016.

\bibitem[{Citron} et~al.(2015){Citron}, {Genda}, and {Ida}]{2015IcarusCitron}
R.~I. {Citron}, H.~{Genda}, and S.~{Ida}.
\newblock {Formation of Phobos and Deimos via a giant impact}.
\newblock \emph{Icarus}, 252:\penalty0 334--338, May 2015.
\newblock arXiv:1503.05623.
\newblock \doi{10.1016/j.icarus.2015.02.011}.

\bibitem[{Collins} et~al.(2004){Collins}, {Melosh}, and
  {Ivanov}]{2004M&PSCollins}
G.~S. {Collins}, H.~J. {Melosh}, and B.~A. {Ivanov}.
\newblock {Modeling damage and deformation in impact simulations}.
\newblock \emph{Meteorit. Planet. Sci.}, 39:\penalty0 217--231, February 2004.
\newblock \doi{10.1111/j.1945-5100.2004.tb00337.x}.

\bibitem[{\'{C}uk} and {Stewart}(2012)]{2012ScienceCukStewart}
M.~{\'{C}uk} and S.~T. {Stewart}.
\newblock {Making the Moon from a Fast-Spinning Earth: A Giant Impact Followed
  by Resonant Despinning}.
\newblock \emph{Science}, 338:\penalty0 1047--, November 2012.
\newblock \doi{10.1126/science.1225542}.

\bibitem[{Frey} and {Schultz}(1988)]{1988GeoRLFreySchultz}
H.~{Frey} and R.~A. {Schultz}.
\newblock {Large impact basins and the mega-impact origin for the crustal
  dichotomy on Mars}.
\newblock \emph{Geophys. Res. Lett.}, 15:\penalty0 229--232, March 1988.
\newblock \doi{10.1029/GL015i003p00229}.

\bibitem[Golabek et~al.(2017)Golabek, Jutzi, Emsenhuber, Gerya, and
  Asphaug]{Companion}
Gregor~J. Golabek, Martin Jutzi, Alexandre Emsenhuber, Taras~V. Gerya, and
  Erik~I. Asphaug.
\newblock {Coupling SPH and thermochemical models of planets: Methodology and example of a Mars-sized body}.
\newblock \emph{Icarus}, 2017.

\bibitem[{Hartmann} and {Davis}(1975)]{1975IcarusHartmannDavis}
W.~K. {Hartmann} and D.~R. {Davis}.
\newblock {Satellite-sized planetesimals and lunar origin}.
\newblock \emph{Icarus}, 24:\penalty0 504--514, April 1975.
\newblock \doi{10.1016/0019-1035(75)90070-6}.

\bibitem[{Hosono} et~al.(2013){Hosono}, {Saitoh}, and
  {Makino}]{2013PASJHosonoSaitohMakino}
N.~{Hosono}, T.~R. {Saitoh}, and J.~{Makino}.
\newblock {Density-Independent Smoothed Particle Hydrodynamics for a Non-Ideal
  Equation of State}.
\newblock \emph{Publ. Astron. Soc. Jpn}, 65:\penalty0 108, October 2013.
\newblock arXiv:1307.0916.
\newblock \doi{10.1093/pasj/65.5.108}.

\bibitem[{Jutzi}(2015)]{2015PSSJutzi}
Martin {Jutzi}.
\newblock {SPH calculations of asteroid disruptions: The role of pressure
  dependent failure models}.
\newblock \emph{Planet. Space Sci.}, 107:\penalty0 3--9, March 2015.
\newblock arXiv:1502.01860.
\newblock \doi{10.1016/j.pss.2014.09.012}.

\bibitem[Jutzi et~al.(2015)Jutzi, Holsapple, Wünneman, and
  Michel]{2015BookJutzi}
Martin Jutzi, K.~Holsapple, K.~Wünneman, and Patrick Michel.
\newblock {Modeling Asteroid Collisions and Impact Processe}.
\newblock In Patrick Michel, Francesca~E. DeMeo, and William~F. Bottke~Jr.,
  editors, \emph{Asteroids IV}, page 679. The University of Arizona Press,
  2015.
\newblock arXiv:1502.01844.

\bibitem[{Marinova} et~al.(2008){Marinova}, {Aharonson}, and
  {Asphaug}]{2008NatureMarinova}
M.~M. {Marinova}, O.~{Aharonson}, and E.~{Asphaug}.
\newblock {Mega-impact formation of the Mars hemispheric dichotomy}.
\newblock \emph{Nature}, 453:\penalty0 1216--1219, June 2008.
\newblock \doi{10.1038/nature07070}.

\bibitem[{Marinova} et~al.(2011){Marinova}, {Aharonson}, and
  {Asphaug}]{2011IcarusMarinova}
M.~M. {Marinova}, O.~{Aharonson}, and E.~{Asphaug}.
\newblock {Geophysical consequences of planetary-scale impacts into a Mars-like
  planet}.
\newblock \emph{Icarus}, 211:\penalty0 960--985, February 2011.
\newblock \doi{10.1016/j.icarus.2010.10.032}.

\bibitem[{Melosh}(1989)]{1989BookMelosh}
H.~J. {Melosh}.
\newblock \emph{{Impact cratering: A geologic process}}.
\newblock Oxford University Press, 1989.

\bibitem[{Melosh}(2007)]{2007M&PSMelosh}
H.~J. {Melosh}.
\newblock {A hydrocode equation of state for SiO2}.
\newblock \emph{Meteorit. Planet. Sci.}, 42:\penalty0 2079--2098, 2007.
\newblock \doi{10.1111/j.1945-5100.2007.tb01009.x}.

\bibitem[{Monaghan}(1992)]{1992ARA&AMonaghan}
J.~J. {Monaghan}.
\newblock {Smoothed particle hydrodynamics}.
\newblock \emph{Annu. Rev. Astron. Astrophys.}, 30:\penalty0 543--574, 1992.
\newblock \doi{10.1146/annurev.aa.30.090192.002551}.

\bibitem[{Monaghan} and {Lattanzio}(1985)]{1985A&AMonaghanLattanzio}
J.~J. {Monaghan} and J.~C. {Lattanzio}.
\newblock {A refined particle method for astrophysical problems}.
\newblock \emph{{Astron. Astrophys.}}, 149:\penalty0 135--143, August 1985.

\bibitem[{Nimmo} et~al.(2008){Nimmo}, {Hart}, {Korycansky}, and
  {Agnor}]{2008NatureNimmo}
F.~{Nimmo}, S.~D. {Hart}, D.~G. {Korycansky}, and C.~B. {Agnor}.
\newblock {Implications of an impact origin for the martian hemispheric
  dichotomy}.
\newblock \emph{Nature}, 453:\penalty0 1220--1223, June 2008.
\newblock \doi{10.1038/nature07025}.

\bibitem[Press(2002)]{NumRecCPP}
William~H. Press.
\newblock \emph{{Numerical recipes in C++ : the art of scientific computing}}.
\newblock Cambridge University Press, 2002.
\newblock {ISBN:~0-521-75033-4}.

\bibitem[{Reinhardt} and {Stadel}(2017)]{2017MNRASReinhardt}
C.~{Reinhardt} and J.~{Stadel}.
\newblock {Numerical aspects of giant impact simulations}.
\newblock \emph{Mon. Not. R. Astron. Soc.}, 467:\penalty0 4252--4263, June
  2017.
\newblock arXiv:1701.08296.
\newblock \doi{10.1093/mnras/stx322}.

\bibitem[{Reufer} et~al.(2012){Reufer}, {Meier}, {Benz}, and
  {Wieler}]{2012IcarusReufer}
A.~{Reufer}, M.~M.~M. {Meier}, W.~{Benz}, and R.~{Wieler}.
\newblock {A hit-and-run giant impact scenario}.
\newblock \emph{Icarus}, 221:\penalty0 296--299, September 2012.
\newblock arXiv:1207.5224.
\newblock \doi{10.1016/j.icarus.2012.07.021}.

\bibitem[{Senft} and {Stewart}(2009)]{2009E&PSLSenftStewart}
L.~E. {Senft} and S.~T. {Stewart}.
\newblock {Dynamic fault weakening and the formation of large impact craters}.
\newblock \emph{Earth Planet. Sci. Lett.}, 287:\penalty0 471--482, October
  2009.
\newblock \doi{10.1016/j.epsl.2009.08.033}.

\bibitem[Speith(2006)]{Habil2006Speith}
Roland Speith.
\newblock {Improvements of the numerical method Smoothed Particle
  Hydrodynamics}.
\newblock Habilitationsschrift, 2006.

\bibitem[Thompson and Lauson(1972)]{ANEOS}
S.~L. Thompson and H.~S. Lauson.
\newblock {Improvements in the CHART-D Radiation-hydrodynamic code III: Revised
  analytic equations of state}.
\newblock Technical Report SC-RR-71 0714, Sandia National Laboratories, 1972.

\bibitem[{Tillotson}(1962)]{Tillotson}
J.~H. {Tillotson}.
\newblock {Metallic Equations of State for Hypervelocity Impact}.
\newblock Technical Report GA-3216, General Atomic, July 1962.

\bibitem[{Wilhelms} and {Squyres}(1984)]{1984NatureWilhelmsSquyres}
D.~E. {Wilhelms} and S.~W. {Squyres}.
\newblock {The martian hemispheric dichotomy may be due to a giant impact}.
\newblock \emph{Nature}, 309:\penalty0 138--140, May 1984.
\newblock \doi{10.1038/309138a0}.

\end{thebibliography}

\end{document}